\documentclass[reprint,superscriptaddress,amsmath,amssymb,aps]{revtex4-2}

\usepackage[utf8]{inputenc}
\usepackage{lmodern}
\usepackage{tabularx}
\usepackage{graphicx}
\usepackage{dcolumn}
\usepackage{bm}
\usepackage{indentfirst}
\usepackage{xcolor}
\usepackage{amsfonts} 
\usepackage{bbold}
\usepackage[version=4]{mhchem}
\usepackage{mathrsfs}
\usepackage{physics}
\usepackage{hyperref}
\hypersetup{colorlinks=true,
            linkcolor=blue,
            filecolor=blue,      
            urlcolor=blue,
            citecolor=blue}

\begin{document}

\title{Quantum gates between distant atoms\\
mediated by a Rydberg excitation antiferromagnet}

\author{Georgios Doultsinos}
\affiliation{Institute of Electronic Structure and Laser, FORTH, 70013 Heraklion, Crete, Greece}
\affiliation{Department of Physics, University of Crete,
Heraklion, Greece}

\author{David Petrosyan}
\affiliation{Institute of Electronic Structure and Laser, FORTH, 70013 Heraklion, Crete, Greece}

\date{\today}

\begin{abstract}
We present a novel protocol for implementing quantum gates between distant atomic qubits connected by an array of neutral atoms that play the role of a quantum bus. The protocol is based on adiabatically transferring the atoms in the array to an antiferromagnetic-like state of Rydberg excitations using chirped laser pulses. 
Upon exciting and de-exciting the atoms in the array under the blockage of nearest neighbors, depending on the state of the two qubits, the system acquires a conditional geometric $\pi$-phase, while the dynamical phase cancels exactly, even when the atomic positions are disordered but nearly frozen in time, which requires sufficiently low temperatures. 
With experimentally relevant parameters, using smooth pulses minimizing the Rydberg-state decay and non-adiabatic errors, we obtain the gate times of $2-3\:\mu$s and gate fidelities of 0.99-0.98 for a pair of atoms separated by $L=20-30\:\mu$m and connected by a quantum bus of several ($3-6$) atoms.  
Optimizing the pulses, we can obtain faster gates exhibiting even better fidelities than those with smooth adiabatic pulses.
\end{abstract}

\maketitle

\section{Introduction}
\label{sec:Intro}

Cold atoms trapped in optical lattices or arrays of microtraps represent a promising platform to realize programmable quantum simulators \cite{bloch2012, Bernien2017, Keesling2019, Omran2019, Bakr2018, Lienhard2018, Scholl2021, ebadi2021quantum, shaw2024, Browaeys2020ManybodyRydberg, morgado2021quantum} and scalable quantum computers \cite{Levine2019Parallel, graham2019, graham2022, evered2023high, ma2023, Tsai2024}. 
Single-qubit initialization, manipulation and read-out with high fidelities in atomic arrays of various geometries have been demonstrated experimentally \cite{xia2015,ma2023}. 
Strong interactions between atoms excited to Rydberg states, i.e., states with high principal quantum number $n$, enable the implementation of two-qubit gates for universal quantum computing \cite{Levine2019Parallel,graham2019,evered2023high,graham2022, ma2023, Tsai2024}. 
Most of these protocols are based on the Rydberg blockade mechanism, whereby resonant laser excitation of one atom suppresses the excitation of other atoms within a distance of several micrometers \cite{Saffman2010InformationRydberg,jaksch2000fast,gaetan2009,Urban2009ObservationBlockade}.
Combined with the optimal control techniques, gate fidelities $F \gtrsim 0.995$ have been experimentally achieved \cite{evered2023high, ma2023, Tsai2024}, approaching the benchmarks set by error correcting codes. 

An important ingredient for efficient quantum information processing on any platform is the ability to realize quantum gates between distant qubits. 
Neutral atoms in Rydberg states interact with each other via the finite-range dipole-dipole or van der Waals interactions, which permit realization of quantum gates between atoms within the blockade distance from each other. 
Quantum gates between distant atomic qubits can be implemented by moving the atoms during the computation \cite{bluvstein2022quantum}, but this is a slow process that can lead to additional motional decoherence. 
Alternatively, optical or microwave photons can be used in hybrid systems as quantum buses to connect static qubits at distant locations \cite{sarkany2015,sarkany2018}. 
But such complicated setups require precise experimental control of many parameters, which is challenging and typically leads to increased loss and decoherence \cite{kaiser2022, ocola2024}. 

Here we suggest a novel protocol to implement quantum gates between distant atoms in an array without actually moving them or interfacing with photons, or performing a sequence of gates between the neighboring atoms \cite{Cesa2007TwoQubitDistantGateLattice,Ahn2022}. 
The crux of our approach is to employ the non-encoding atoms of the array as a quantum bus that connects distant qubits. Our protocol is based on simultaneously driving the qubit and bus atoms with a global chirped laser pulse which induces adiabatic transition between the uniform ground (ferromagnetic) state with no excitations and an antiferromagnetic-like (AFM-like) state of Rydberg excitations \cite{pohl2010dynamical, Tzortzakakis2022Microscopic, Keesling2019, Bakr2018, Lienhard2018, Scholl2021,  ebadi2021quantum}. 
Depending on the state of the qubits connected to the bus, the system reaches the many-body state with different number $\nu_r$ of Rydberg excitations acquiring a dynamical phase and a geometric phase $\pi \nu_r \!\! \mod(2\pi)$ that depends only on the parity of $\nu_r$. 
After the system is transferred back to the ground state using an identical chirped pulse, the dynamical phase is exactly canceled but the geometric phase remains, and the total transformation is equivalent to a \textsc{cz} phase gate between the distant qubits.

Before continuing, we note the previous relevant work. 
Adiabatic rapid transfer with global laser pulses has been used to implement two- and multi-qubit gates with atoms within the Rydberg blockade distance \cite{Saffman2020SymmetricCZ, pelegri2022high}, and three-qubit gates for atoms with strong nearest-neighbor interaction have also been developed \cite{Levine2019Parallel, jandura2022time}.
Another relevant work is a proposal to implement the \textsc{cz} gate between distant atomic qubits connected to a chain of mediating atoms \cite{weimer2012long}. 
In this scheme, the many atoms in the chain can be approximated by a continuous one-dimensional system which is transferred to a crystalline state of Rydberg excitations that repel each other and thereby imprint dynamical phases onto the end ground-Rydberg qubits. 
%Hence, to achieve good fidelities, the dynamics of the system and the accumulated dynamical phase should be controlled with high precision. 
In contrast, in our protocol the interaction-induced dynamical phases are canceled and the remaining geometric phases depend only on the parity of the intermediate state with multiple Rydberg excitations of atoms in a discrete lattice. 
Our protocol thus requires only a few atoms to operate while being immune to uncertainties in interatomic distances. But any preexisting disorder of the array must be ``frozen" during the execution of the protocol, which means that the characteristic time of thermal motions must be much longer than the operation time of the gate. 
This can be achieved with sufficiently cold atoms commonly used in Rydberg atom quantum computers and simulators \cite{Browaeys2020ManybodyRydberg, morgado2021quantum, ebadi2021quantum, shaw2024, Levine2019Parallel, graham2019,evered2023high, graham2022, ma2023, Tsai2024}. 

The paper is organized as follows.
In Sec.~\ref{sec:PXPandvdW} we introduce the system of laser-driven atoms in a lattice and explain the principles of operation of the quantum gate using first a simple $PXP$ model and then the complete model including finite-strength, long-range van der Waals interactions between the atoms. 
In Sec.~\ref{sec:gateFE} we analyze the resulting gate fidelity numerically and analytically.
Our conclusions are summarized in Sec.~\ref{sec:conclud}.
Details of calculations of the many-body spectrum and eigenstates of the system, dynamical and geometric phases, parity errors, and change of the sign of interaction between the atoms are presented in Appendices~\ref{app:HSpctrSym}-\ref{app:BtomB}.

\section{Interacting many-body system}
\label{sec:PXPandvdW}

%%%%%%%%%%%%%%%%%%%%%%%%%%%%%%%%%%%%%%%%%%%%%%%%
\begin{figure}[t]
\includegraphics{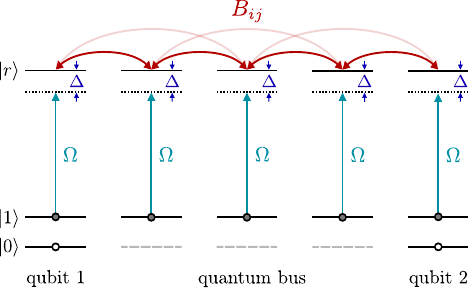}
\caption{Schematics of a chain of $N$ atoms with the ground state sublevels $\ket{0,1}$ and the Rydberg state $\ket{r}$ coupled to state $\ket{1}$ by a global laser field with the Rabi frequency $\Omega$ and detuning $\Delta$. Atoms in the Rydberg state $\ket{r}$ interact via the pairwise long-range interaction $B_{ij}$. 
The first and last atoms of the chain represent qubits, while the remaining $N-2$ atoms play the role of a quantum bus connecting the qubits.}
\label{fig:als}
\end{figure}
%%%%%%%%%%%%%%%%%%%%%%%%%%%%%%%%%%%%%%%%%%%%%%%%

We consider a chain of $N$ neutral atoms trapped by optical tweezers at equidistant positions. 
The relevant states of the atoms are a pair of long-lived hyperfine sub-levels $\ket{0}$ and $\ket{1}$ of their ground state and a highly excited Rydberg state $\ket{r}$, see Fig.~\ref{fig:als}. 
The first and the last atoms of the chain can store qubits as superpositions of their states $\ket{0}$ and $\ket{1}$, while the remaining $N-2$ atoms are prepared in state $\ket{1}$ to serve as a quantum bus connecting the qubits.
%The qubit states can be initialized and manipulated with high fidelity using microwave pulses and spatially-inhomogeneous magnetic fields or non-resonant lasers that induce different Zeeman or AC Stark level-shifts of individual atoms. 
State $\ket{1}$ of each atom is coupled by a common laser field to the Rydberg state $\ket{r}$ with the time-dependent Rabi frequency $\Omega(t)$ and detuning $\Delta(t)$, while state $\ket{0}$ remains passive. In the frame rotating with the laser frequency, the atom-field interaction is described by the Hamiltonian ($\hbar=1$)
\begin{equation}\label{eq:AFHam}
    \mathcal{H}_{\mathrm{af}}(t) = \tfrac{1}{2} \Omega(t)\sum_{i=1}^{N} \ket{r_i}\bra{1_i} + \mathrm{H.c.} -\Delta(t)\sum_{i=1}^{N} \ket{r_i}\bra{r_i} .  
\end{equation}
Atoms excited to Rydberg states interact via long-range pairwise interactions described by  
\begin{equation}\label{eq:AAHam}
\mathcal{H}_{\mathrm{aa}} = \sum_{i>j}B_{ij}\ket{r_ir_j}\bra{r_ir_j} .
\end{equation}
We assume van der Waals interaction $B_{ij} = \tfrac{B}{|i - j|^6}$, where $B = C_6/a^6$ is the dominant interaction between the neighboring atoms expressed through the van der Waals coefficient $C_6$ and the lattice spacing $a$. 
To ensure the Rydberg blockade, we require that $B_{i,i+1} = B \gg \max(\Delta,\Omega)$ which suppresses simultaneous excitation of neighboring atoms. 
At the same time, we assume $B_{i,i+2} = B/2^6 < \max ( \Omega)$, so that the atoms separated by two (or more) lattice sites interact only weakly and their simultaneous excitation is not suppressed.

\subsection{Effective \textit{PXP} model}
\label{subsec:PXP}

\begin{figure*}[!t]
\includegraphics{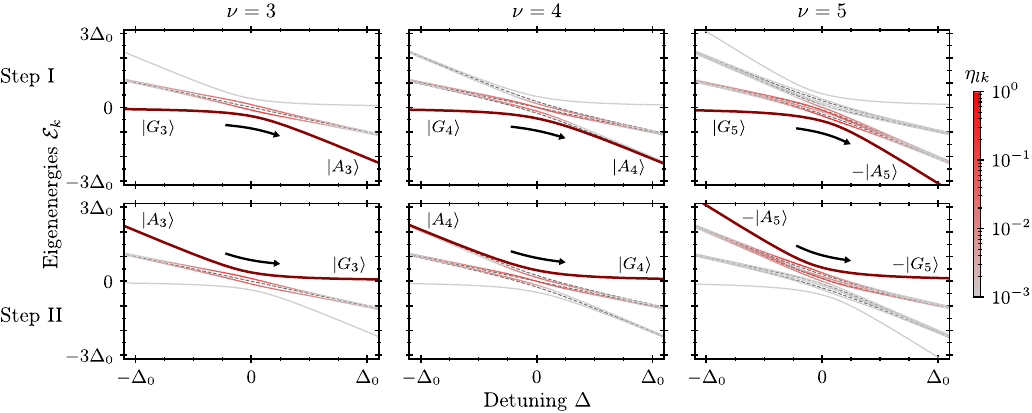}
 \caption{Energy eigenvalues $\mathcal{E}_k$ of the effective Hamiltonian $\mathcal{H}_{\mathrm{PXP}}$ for $\nu = 3, 4, 5$ atoms (left, center, right columns) vs  detuning $\Delta$ in the range $-\Delta_0 < \Delta < \Delta_0$, $\Delta_0 = 3\Omega$.
 In the first step (upper panels), we adiabatically follow the lowest-energy $\mathcal{E}_1$ eigenstate $\ket{\alpha_1}$ (thick brown line) asymptotically connected to $\ket{G_\nu}$ for $\Delta < -\Delta_0$ and to $(-1)^{\nu_r}\ket{A_\nu}$ for $\Delta > \Delta_0$ with $\nu_r = \lceil \nu/2 \rceil$ Rydberg excitations.
 In the second step (lower panels), we adiabatically follow the highest-energy $\mathcal{E}_m$ eigenstate $\ket{\alpha_m}$ (thick brown line) asymptotically connected to $\ket{A_\nu}$ for $\Delta < -\Delta_0$ and to $\ket{G_\nu}$ for $\Delta > \Delta_0$. 
 Non-adiabatic transition probabilities from states 
 $\ket{\alpha_{l=1,m}}$ to other instantaneous eigenstates $\ket{\alpha_k}$ are quantified by the dimensionless parameter $\eta_{l k} = |\bra{\alpha_{l}}\partial_t\ket{\alpha_k}|^2\tau/\Delta_0$ (red color depths of the solid lines). 
 Antisymmetric eigenstates (gray dashed lines) are ``dark'' for all $\Delta$ as they remain decoupled from the laser acting symmetrically on all the atoms. 
 For odd number of atoms ($\nu=3,5$), there is a finite energy gap in the vicinity of $\Delta =0$ which protects 
 the adiabatic evolution of $\ket{\alpha_{1,m}}$. For even number of atoms ($\nu=4$), there are states with energies approaching $\mathcal{E}_{1,m}$ for $|\Delta| \gg \Omega$, but their non-adiabatic coupling to $\ket{\alpha_{1,m}}$ decreases even more rapidly.}
    \label{fig:PXPspectrum}
\end{figure*}

Our gate protocol relies on the adiabatic preparation of AFM-like state of Rydberg excitations in a one-dimensional lattice using frequency-chirped laser pulses 
\cite{pohl2010dynamical, Tzortzakakis2022Microscopic, Bernien2017, Keesling2019, Bakr2018, Lienhard2018, Scholl2021,  ebadi2021quantum,  ebadi2022quantum}. 
To understand the main idea, it is instructive to consider first an effective model \cite{Lesanovsky2012FibonacciAnyons, Turner2018pxpScars} that forbids simultaneous excitation of neighboring atoms and neglects long-range interactions, $B_{i,i+j} = 0 \, \forall \, j \geq 2$; later we will consider the corrections to this model and discuss the implications. 
Assuming $\nu$ atoms with states $\ket{1},\ket{r}$, the effective ($PXP$) Hamiltonian  is 
\begin{equation}\label{eqs:PXPhamiltonian}
\mathcal{H}_{PXP}(t) = \tfrac{1}{2}\Omega(t)\sum_{i=1}^\nu P_{i-1} X_i P_{i+1} -\Delta(t)\sum_{i=1}^\nu \ket{r_i}\bra{r_i} ,
\end{equation}
where $X_i = \ket{r_i}\bra{1_i} + \ket{1_i}\bra{r_i}$ are the transition operators and $P_i = \ket{1_i}\bra{1_i}$ are the projection operators for atoms $i=1,2, \ldots , \nu$, while $P_0 = P_{\nu+1} \equiv 1$ as we assume open boundaries. 
For any $\Omega, \Delta$, the instantaneous eigenstates $\ket{\alpha_k}$ of $\mathcal{H}_{PXP}$ are defined via 
\[
   \mathcal{H}_{PXP} \ket{\alpha_k} =  \mathcal{E}_k  \ket{\alpha_k} ,
\]
where $\mathcal{E}_k$ are the corresponding eigenenergies ordered as $\mathcal{E}_1 \leq \mathcal{E}_2 \leq \ldots \leq \mathcal{E}_m$.
The number of eigenstates, and the size of the corresponding Hilbert space, is $m=F_{\nu +2}$, where $F_n$ are the Fibonacci numbers \cite{Lesanovsky2012FibonacciAnyons}.
The spectrum of $\mathcal{H}_{PXP}$ has the symmetry property $\mathcal{E}_k(\Omega,\Delta)= -\mathcal{E}_{m-k+1}(\Omega,-\Delta)$, see Fig.~\ref{fig:PXPspectrum} and Appendix~\ref{app:HSpctrSym}.

% %%%%%%%%%%%%%%%%%%%%%%%%%%%%%%%%%%%%%%%%%%%%%%%
% %%%%%%%%%%%%%%%%%%%%%%%%%%%%%%%%%%%%%%%%%%%%%%%

Consider a chain of $\nu$ ($=N-2,N-1$ or $N$) atoms initially all in state $\ket{1}$. 
For a large negative detuning, $\Delta <0 $ and  $-\Delta \gg |\Omega|$, the initial state of the system $\ket{G_{\nu}} \equiv \ket{11\ldots1}$ coincides with the lowest energy eigenstate $\ket{\alpha_1}$ with $\mathcal{E}_1 \simeq 0$, while for a large positive detuning $\Delta \gg |\Omega|$, the lowest energy eigenstate $\ket{\alpha_1}$ with $\mathcal{E}_1 \simeq -\nu_r \Delta$ corresponds to the AFM-like state  $\ket{A_\nu}$ with $\nu_r = \lceil \nu/2 \rceil$ Rydberg excitations (see Fig.~\ref{fig:PXPspectrum} upper panels).
More precisely, for odd $\nu$, we have a single ordered state $\ket{A_\nu} = \ket{r1r1\ldots 1r}$ with $\nu_r = (\nu+1)/2$ Rydberg excitations; while 
for even $\nu$, the lowest-energy state $\ket{A_\nu}$ with $\nu_r = \nu/2$ Rydberg excitations consists of a superposition of two ordered configurations $\ket{r1r1 \ldots r1}$ and $\ket{1r1r \ldots 1r}$ and $(\nu-2)/2$ configurations of the form $\ket{r1\ldots r11r \ldots 1r}$ with the excited boundaries and one defect at various positions in the bulk (see Appendix~\ref{app:AFMstatePXP}). 
Hence, starting with state $\ket{G_{\nu}}$, by turning on the Rabi frequency of the laser field $\Omega(t)$ during time $\tau$ while sweeping the detuning $\Delta(t)$ from some large negative value $\Delta(0) = -\Delta_0$ to the large positive value $\Delta(\tau) = \Delta_0 \gg \max(\Omega)$, we adiabatically follow the instantaneous ground state of the system $\ket{\alpha_1}$ which connects state $\ket{G_{\nu}}$ to state $e^{i\phi} \ket{A_\nu} = e^{i\phi_d^{(1)}} (-1)^{\nu_r} \ket{A_\nu} $, where the total phase $\phi = \phi_d^{(1)} + \phi_g$ contains the dynamical phase $\phi_d^{(1)} = \int_0^\tau \mathcal{E}_1(t) dt$ and the geometric phase $\phi_g = \pi \nu_r$ that depends on the parity of $\nu_r$ (see Appendix~\ref{app:parityphasePXP}). 

Next, to transfer the system from the AFM-like state $\ket{A_\nu}$ back to the ground state $\ket{G_\nu}$, we repeat the laser pulse. Now at $\Delta = -\Delta_0$ state $\ket{A_\nu}$ corresponds to the eigenstate $\ket{\alpha_m}$ with the highest energy $\mathcal{E}_m \simeq \nu_r \Delta_0$ (see Fig.~\ref{fig:PXPspectrum} lower panels).
%in the low-energy band of the system, $\mathcal{E}_k \in [0,\nu_r \Delta_0] \; \forall \, k \in [1,m]$. 
By turning on $\Omega(t>\tau)$ for time $\tau$ while sweeping the detuning $\Delta(t>\tau)$ from $-\Delta_0$ to $\Delta_0$, we adiabatically follow the instantaneous eigenstate $\ket{\alpha_m}$ that connects state $\ket{A_\nu}$ at $\Delta=-\Delta_0$ to the state $e^{i\phi_d^{(m)}} \ket{G_\nu}$ at $\Delta=\Delta_0$. The accumulated dynamical phase $\phi_d^{(m)} = \int_\tau^{2\tau} \mathcal{E}_m(t) dt = -\phi_d^{(1)}$ has exactly the same absolute value as $\phi_d^{(1)}$ but opposite sign, since $\mathcal{E}_1 (t) = -\mathcal{E}_m (2\tau - t)$ and we assume linear sweep of $\Delta(t)$. Hence, after two applications of the laser pulses, at time $t=2\tau$ the dynamical phase is canceled and the state of the system is $(-1)^{\nu_r} \ket{G_{\nu}}$ with $\nu_r = (\nu+1)/2$ for $\nu$ odd, and $\nu_r = \nu/2$ for $\nu$ even.   

\subsubsection{Gate protocol}
\label{subsubsec:GateProtocol}

Recall now the truth table of the \textsc{cz} gate for two qubits: 
\[
\ket{q_1} \ket{q_2} \overset{\textsc{cz}}{\longrightarrow} (-1)^{q_1q_2} \ket{q_1} \ket{q_2} \quad \mathrm{for} \quad q_{1,2}=0,1 ,
\]
i.e., only state $\ket{1} \ket{1}$ acquires the sign change $(-1)$ (or phase shift $\pi$), while the other states $\ket{0} \ket{0},\ket{0} \ket{1},\ket{1} \ket{0}$ remain unchanged.
Our qubits are represented by the two end atoms in a chain of $N$ atoms, with the intermediate $N-2$ atoms prepared in state $\ket{1}$.  
Assume $N$ is odd. Then, if both qubits 1 and 2 are in state $\ket{0}$, we have $\nu=N-2$ and $\nu_r=(N-1)/2$; and if one of the qubits is in state $\ket{0}$ and the other in state $\ket{1}$, we have $\nu=N-1$ and again $\nu_r=(N-1)/2$; while if both qubits are in state $\ket{1}$, we have $\nu=N$ and $\nu_r=(N+1)/2 = (N-1)/2 + 1$. 
Hence, our protocol realizes the transformation $\ket{q_1} \ket{11 \ldots 1} \ket{q_2} \rightarrow (-1)^{(N-1)/2 + q_1q_2} \ket{q_1} \ket{11 \ldots 1} \ket{q_2}$ which, upon factoring out the common for all the inputs factor $(-1)^{(N-1)/2}$ and the intermediate quantum bus in state $\ket{11 \ldots 1}$, is precisely the \textsc{cz} gate for the two qubit. 
Table~\ref{tab:InputsConfigsParity} illustrates the transformation for $N=5$.

%%%%%%%%%%%%%%%%%%%%%%%%%%%%%%%%%%%%%%%%%%%%%%%%%%
\begin{table}
\caption{Input states $\ket{q_1 q_2}$ of the two qubits connected to the two ends of the bus of $N-2$ atoms correspond to initial configurations with $\nu = N-2,N-1,$ or $N$ atoms in state $\ket{1}$.
Transfer to the AFM-like configurations with the corresponding parity $(-1)^{\nu_r}$ and back leads to the \textsc{cz} gate. 
Atoms in states $\ket{0},\ket{1}$ and $\ket{r}$ are denoted by open circles, filled black circles and filled red circles, respectively.}
\label{tab:InputsConfigsParity}
\centering
\begin{tabularx}{\columnwidth}{ 
   >{\centering\arraybackslash}X 
   >{\centering\arraybackslash}X  
   >{\centering\arraybackslash}X  
   >{\centering\arraybackslash}X 
   >{\centering\arraybackslash}X}
    \hline
    \hline
    $\ket{q_1q_2}$ & Initial &$\rightleftharpoons$&  AFM-like  & Parity\\
    &configuration&&configuration&\\
    \hline
    $\ket{00}$ & $\circ\bullet\bullet\bullet\circ$ &&
    $\circ\color{red}\bullet\color{black}\bullet\color{red}\bullet\color{black}\circ$& $+1$\\ 
    $\ket{01}$ & $\circ\bullet\bullet\bullet\bullet$ &&  
                 $\circ\color{red}\bullet\color{black}\bullet\bullet\color{red}\bullet\color{black}$& $+1$\\
    $\ket{10}$ & $\bullet\bullet\bullet\bullet\circ$ && 
                 $\color{red}\bullet\color{black}\bullet\bullet\color{red}\bullet\color{black}\circ$ &$+1$\\
    $\ket{11}$ & $\bullet\bullet\bullet\bullet\bullet$ &&${\color{red}\bullet}\bullet{\color{red}\bullet}\bullet{\color{red}\bullet}$& $-1$ \\
    \hline
    \hline  
\end{tabularx}
\end{table}
%%%%%%%%%%%%%%%%%%%%%%%%%%%%%%%%%%%%%%%%%%%%%%%%%%

Similarly for even $N$: 
For input state $\ket{0}\ket{0}$ we have $\nu_r=(N-2)/2 =N/2-1$, while for states $\ket{0}\ket{1}$, $\ket{1}\ket{0}$ and $\ket{1}\ket{1}$ we have
$\nu_r=N/2$, and therefore 
$\ket{q_1} \ket{11 \ldots 1} \ket{q_2} \rightarrow (-1)^{N/2 - \bar{q}_1 \bar{q}_2} \ket{q_1} \ket{11 \ldots 1} \ket{q_2}$, where $\bar{q}_{1,2} \equiv 1-q_{1,2}$. 
Factoring out the state of the bus and the common phase factor $(-1)^{N/2}$, we obtain that the input state $\ket{0}\ket{0}$ acquires the sign change $(-1)$, which is equivalent to the \textsc{cz} gate up to the qubit flips $\ket{0} \leftrightarrow \ket{1}$ realized by the single-qubit $X$ gates.

% %%%%%%%%%%%%%%%%%%%%%%%%%%%%%%%%%%%%%%%%%%%%%%%
\begin{figure}[t]
\includegraphics[width=\columnwidth]{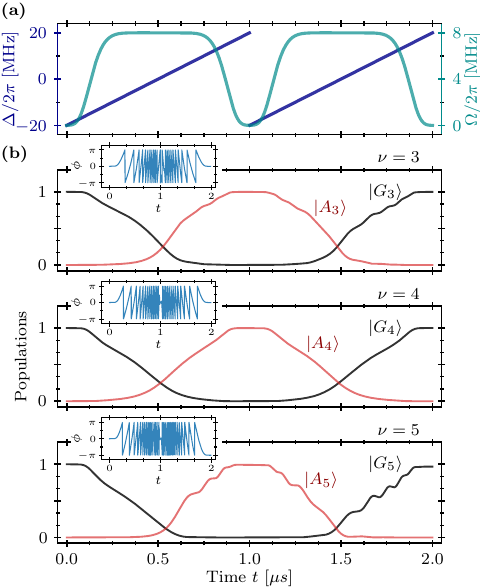}
    \caption{(a) Time dependence of the Rabi frequency $\Omega(t)$ and detuning $\Delta(t)$ of the global laser field acting on the transition $\ket{1} \leftrightarrow \ket{r}$ of all the atom in the chain. 
    Two identical chirped pulses of duration $\tau=1\:\mu\mathrm{s}$ and $\max |\Omega| = \Omega_0 = 2\pi \times 8 \:\mathrm{MHz}$, $\max |\Delta| = \Delta_0= 2\pi \times 20 \:\mathrm{MHz}$ transfer $\nu$ atoms from the collective ground state $\ket{G_\nu}$ to the AFM-like state $\ket{A_\nu}$ and back. 
    (b)~Dynamics of populations of states $\ket{G_\nu}$ and $\ket{A_\nu}$ for the chains of $\nu= 3, 4, 5$ atoms (upper, middle, lower panels) governed by the effective Hamiltonian $\mathcal{H}_{PXP}$. Insets show the total phase $\phi(t) = \arg\bra{\Psi(0)}\ket{\Psi(t)}$ acquired by the state of the system $\ket{\Psi(t)}$ during the evolution; The final phase $\phi(2\tau) = 0$ for $\nu=3,4$ and $\phi(2\tau) = \pi$ for $\nu=5$.}
    \label{fig:PulsesPopulations}
\end{figure}
% %%%%%%%%%%%%%%%%%%%%%%%%%%%%%%%%%%%%%%%%%%%%%%%

\subsubsection{Dynamics of the system}
\label{subsubsec:PXPDynamics} 

We now examine the dynamics of the system more quantitatively. 
During the first step, $0 < t \leq \tau$, a chirped laser pulse of duration $\tau$ with amplitude and frequency detuning given by 
\begin{subequations}\label{eqs:PulseProfile}
    \begin{align}
        \Omega(t) &=\Omega_0 \frac{e^{-(t-\tau/2)^8/\sigma^8} - e^{-(\tau/2\sigma)^8}}{1-e^{-(\tau/2 \sigma)^8}}, \\
        \Delta(t) &= \beta \, (t -\tau/2) ,
    \end{align}
\end{subequations}
with $\sigma \simeq 0.385 \tau$ and $\beta = 2 \Delta_0/\tau$, transfers $\nu$ atoms initially in state $\ket{1}$ from the collective ground state $\ket{G_\nu}$ to the AFM-like state $\ket{A_\nu}$ with $\nu_r = \lceil \nu/2 \rceil$ Rydberg excitations.
During the second step, an identical pulse shifted in time by $\tau$, $t \to t+\tau$, transfers the atoms from $\ket{A_\nu}$ back to $\ket{G_\nu}$, see Fig.~\ref{fig:PulsesPopulations}. 
During the evolution, the state of the system $\ket{\Psi(t)}$ follows the adiabatic eigenstates of $\mathcal{H}_{PXP}$, $\ket{\alpha_1}$ for $0 < t \leq \tau$ and $\ket{\alpha_m}$ for $\tau < t \leq 2\tau$, and acquires the total phase $\phi(t) = \arg\bra{\Psi(0)}\ket{\Psi(t)}$ which includes the dynamical phase $\phi_\mathrm{d}(t) = \int_0^{t} \langle \mathcal{H}\rangle dt'$ and geometric phase $\phi_\mathrm{g} = \phi(t) - \phi_d(t)$. 
Due to the symmetry of the adiabatic energy eigenvalues, $\mathcal{E}_1 (t) = -\mathcal{E}_m (2\tau - t)$, the dynamical phase is canceled at the end of the process, $t=2\tau$, and only the geometric phase $\phi_g = \nu_r \pi \! \mod(2\pi)$, that depends on the parity of $\nu_r$, remains. 

According to the adiabatic theorem \cite{MessiahQM1961vII}, in order for the system to adiabatically follow an instantaneous eigenstate $\ket{\alpha_{l}(t)}$ of a time-dependent Hamiltonian $\mathcal{H}(t)$, the non-adiabatic transition rates $\bra{\alpha_{l}(t)}\partial_t\ket{\alpha_k(t)}$ to all the other instantaneous eigenstates $\ket{\alpha_k(t)}$ should be small compared to their energy separations $\mathcal{E}_{lk}(t) \equiv | \mathcal{E}_l(t) - \mathcal{E}_k(t)|$, 
\[
\left| \frac{\bra{\alpha_{l}}\partial_t\ket{\alpha_k}}{\mathcal{E}_l - \mathcal{E}_k } \right| \ll 1 .
\]

In Fig.~\ref{fig:PXPspectrum}, we quantify the non-adiabatic transition rates from $\ket{\alpha_{l=1,m}}$ by the dimensionless parameter $\eta_{l k} = |\bra{\alpha_{l}}\partial_t\ket{\alpha_k}|^2\tau/\Delta_0$. 
Some of the eigenstates $\ket{\alpha_{k_{\mathrm{as}}}}$ of $\mathcal{H}_{PXP}$ are antisymmetric with respect to the spatial inversion operator 
$\mathcal{I}=\sum_{i=1}^{\lfloor\nu/2\rfloor}\ket{1_i 1_{\nu-i+1}}\bra{1_i 1_{\nu-i+1}} + \ket{1_i r_{\nu-i+1}}\bra{r_i 1_{\nu-i+1}} + \ket{r_i 1_{\nu-i+1}}\bra{1_i r_{\nu-i+1}} + \ket{r_i r_{\nu-i+1}}\bra{r_i r_{\nu-i+1}}$, 
which is a constant of motion. 
Hence, if the laser drives symmetrically all the atoms and decay and dephasing are small, the antisymmetric states remain ``dark'', or decoupled $\eta_{l k_\mathrm{as}} =0$, for any rate of change of the laser parameters -- here $\Delta(t)$ -- and independent on $\mathcal{E}_{lk_\mathrm{as}}$. 
The remaining eigenstates $\ket{\alpha_k}$ are in general ``bright'', with transition rates $\eta_{l k} \neq 0$ varying with $\Delta$. 

We observe in Fig.~\ref{fig:PXPspectrum} that for odd $\nu$, there is a finite energy gap $\mathcal{E}_{12} = |\mathcal{E}_{1} - \mathcal{E}_{2}|$ between the lowest energy eigenstate $\ket{\alpha_1}$ and the first excited state $\ket{\alpha_2}$ that tends asymptotically (for $\Delta \gg |\Omega|$) to a configuration with one less Rydberg excitation and energy $\mathcal{E}_2 \to -(\nu_r-1) \Delta$.  
Thus, during the first pulse, the lowest energy eigenstate $\ket{\alpha_1}$ is protected by this energy gap that attains the minimal value $\delta \mathcal{E} \equiv \min [\mathcal{E}_{12}] \sim \Omega_0/\nu$ in the vicinity of $\Delta = 0$. 
The system with odd number of atoms $\nu$ can then be described by an effective Landau-Zener theory to estimate the preparation fidelity of the AFM-like state $\ket{A_\nu}$ at time $t=\tau$ when $\Delta = \Delta_0 \gg |\Omega|$ \cite{Tzortzakakis2022Microscopic}. 

For even $\nu$, we observe that in the vicinity of $\Delta = 0$ there is too a finite energy gap between the lowest energy eigenstate $\ket{\alpha_1}$ and other eigenstates $\ket{\alpha_{k>1}}$. 
Some of these states likewise tend asymptotically (for $\Delta \gg |\Omega|$) to configurations with one less Rydberg excitation and energy $\mathcal{E}_k \to -(\nu_r-1) \Delta$. 
But there are also $\nu/2$ eigenstates $\ket{\alpha_k}$ ($k=2,3, \ldots, \nu/2+1$) with energies asymptotically (for $\Delta \gg |\Omega|$) approaching the ground state energy $\mathcal{E}_1 \to -\nu_r\Delta$.
These eigenstates correspond to different superpositions of the AFM-like configurations, $\ket{\alpha_k} \simeq \ket{\aleph_k}$, as detailed in Appendix~\ref{app:AFMstatePXP}. 
But while the energy gap $\mathcal{E}_{1k}$ between the real ground state $\ket{\alpha_1}$ and other AFM-like states $\ket{\alpha_{k}}$ ($1<k \leq \nu/2+1$) decreases with increasing $\Delta$, their non-adiabatic transition rates $\bra{\alpha_{1}}\partial_t\ket{\alpha_k}$ decrease even faster, partially suppressing the transitions away from the ground state. 
The relevant energy gap $\delta \mathcal{E} = \min [\mathcal{E}_{1k}]$ is now between the state $\ket{\alpha_1}$ and the state $\ket{\alpha_k}$ ($k=\nu/2+2$)  that tends asymptotically (for $\Delta \gg |\Omega|$) to a configuration with one less Rydberg excitation and energy $\mathcal{E}_k \to -(\nu_r-1) \Delta$, as detailed in Sec.~\ref{subsec:decayLZ} and Appendix~\ref{app:Eleakagewp}.   

During the second pulse, exactly the same arguments apply for the dynamics of the system initially in state $\ket{A_\nu}$ corresponding to the highest energy eigenstate $\ket{\alpha_m}$. 
Again, for odd $\nu$, the adiabatic evolution of $\ket{\alpha_m}$ is protected by the finite energy gap $\delta \mathcal{E} = \min [\mathcal{E}_{m,m-1}] \sim \Omega_0/\nu$. 
But for even $\nu$, we now start at $t =\tau +0$ from a nearly degenerate subspace of AFM-like states $\ket{\alpha_{k'}}$ ($k'=m,m-1,\ldots,m-\nu/2$) all having $\nu_r=\nu/2$ Rydberg excitations. 
The important observation is that, even if at the end of the first pulse we populate not only the lowest energy state $\ket{\alpha_1}$ but also other nearly-degenerate AFM-like states $\ket{\alpha_{k}}$ ($k=1,2, \ldots, \nu/2+1$) due to non-adiabatic transitions between them, during the second pulse the corresponding AFM-like states $\ket{\alpha_{k'}}$ ($k'=m, m-1, \ldots,m-\nu/2$) in the highest energy manifold would undergo, up to the parity $(-1)^{\nu_r}$ (see Appendix~\ref{app:parityphasePXP}), the time-reversed dynamics and end up in state $\ket{G_\nu}$ for $\Delta = \Delta_0$ at $t=2\tau$, 
provided the temporal profiles of the first and the second pulses are related as $\Omega(t)=\Omega(2\tau-t)$ and $\Delta(t) = -\Delta(2\tau-t)$.  
Similarly to the first step, only the transitions to the states that tend asymptotically (for $\Delta \gg |\Omega|$) to a configuration with one Rydberg excitation and energy $\mathcal{E}_k \to - \Delta$ result in error, see Sec.~\ref{subsec:decayLZ} and Appendix~\ref{app:Eleakagewp}.

\subsection{van der Waals interacting model}
\label{subsec:vdW}

Consider now the full model for the van der Waals interacting atoms, described by the Hamiltonian $\mathcal{H} = \mathcal{H}_{\mathrm{af}} + \mathcal{H}_{\mathrm{aa}}$. 
Since we assumed $|B| \gg \Delta_0,\Omega_0$, simultaneous excitation of neighboring atoms is suppressed and the effective $PXP$ model in Eq.~(\ref{eqs:PXPhamiltonian}) still captures the main properties of the system. 
But there are now corrections due to the long range and finite strength of the interatomic interactions $B_{ij}$ in Eq.~(\ref{eq:AAHam}). 
The dominant long-range interaction corresponds to the interaction between the next nearest-neighbor atoms described by 
\begin{equation}
\delta \mathcal{H}_\mathrm{aa} = B_2 \sum_{i=1}^{\nu-2}Q_i Q_{i+2} , \label{eq:LRAAHam}
\end{equation}
where $B_2 = B_{i,i+2} = B/2^6$ is the interaction strength and $Q_i = \ket{r_i}\bra{r_i}$ are projectors onto the Rydberg states of the atoms. 
The finite strength of the nearest-neighbor interaction $B$ leads to incomplete blockade and virtual Rydberg excitation of an atom next to the already excited atom. 
The corresponding corrections are obtained via the second order Schrieffer-Wolff transformation \cite{DiVincenzo2011Schieffer} leading to 
\begin{align}
    \delta \mathcal{H}_S = 
    &-S_B \sum_{i=1}^{\nu} P_{i-1}P_iQ_{i+1} + Q_{i-1}P_iP_{i+1} \nonumber \\   
    &-S_{2B} \sum_{i=1}^{\nu} Q_{i-1}P_iQ_{i+1} , \label{eq:HSshift}
\end{align}
where $S_B = \frac{|\Omega|^2}{4(B - \Delta)}$ and $S_{2B} = \frac{|\Omega|^2}{4(2B - \Delta)}$ are the second-order level shifts of an atom in state $\ket{1_i}$ having one or two neighbors in Rydberg state, while $P_0 = P_{\nu+1} \equiv 1$ and $Q_0 = Q_{\nu+1} \equiv 0$ for open boundaries.
In addition, there is second-order hopping of Rydberg excitations between the neighboring lattice sites described by
\begin{equation}
    \delta \mathcal{H}_\mathrm{h} = -S_B \sum_{i=1}^{\nu-2} P_{i-1} (\sigma_{i+1}^+\sigma_{i}^- + \sigma_{i}^+\sigma_{i+1}^-) P_{i+2} , \label{eq:HShop}
\end{equation}
where $\sigma_i^+ = \ket{r_i}\bra{1_i}$ and $\sigma_i^- = \ket{1_i}\bra{r_i}$ are transition operators. 
Closely related physics was discussed in \cite{Molmer2024AnisotropicHeisenberg}.

\subsubsection{Energy spectrum}

For odd $\nu$, the spectrum of the adiabatic eigenstates of the full Hamiltonian, $\mathcal{H} \ket{\alpha_k} = \mathcal{E}_k \ket{\alpha_k}$, has the same form as for the $PXP$ model (see Fig.~\ref{fig:PXPspectrum} for $\nu = 3,5$) but with important differences. 
In particular, the energy of the AFM-like state $\ket{A_\nu} = \ket{r1r1 \ldots 1r}$ with $\nu_r = (\nu +1)/2$ Rydberg excitations is shifted by $B_2 (\nu_r - 1)$ due to the long-range interactions of Eq.~(\ref{eq:LRAAHam}).
In addition, there is a second-order energy shift $S_{2B} (\nu_r - 1)$ due to virtual Rydberg excitations of $(\nu_r - 1)$ incompletely blockaded atoms in state $\ket{1}$ as per the second term in Eq.~(\ref{eq:HSshift}). 
Thus the lowest energy AFM state $\ket{\alpha_1} = \ket{A_\nu}$ for $\Delta \gg  |\Omega|$ and the highest energy AFM state $\ket{\alpha_m} = \ket{A_\nu}$ for $-\Delta \gg  |\Omega|$ have the energies
\[
\mathcal{E}_{1,m} = -\nu_r ( \Delta + S ) + (\nu_r-1) (B_2 -S_{2B}) \;\; \mathrm{for} \;\; \Delta \gtrless 0 ,  
\]
where $S=|\Omega|^2/4\Delta$, see Appendix~\ref{app:AFMstatevdW}. 
Note that $S_{2B}$ is a function of detuning $\Delta$ and, for a fixed $B \gg |\Delta|$, it has different values for $\Delta>0$ and $\Delta<0$.
Hence, the long range and finite strengths of the interatomic interactions $B_{ij}$ break the symmetry 
$\mathcal{E}_1 (\Omega,\Delta,B) \neq - \mathcal{E}_m (\Omega,-\Delta,B)$. 
But this symmetry is recovered if we change the sign of both $B$ and $\Delta$ (see Appendix~\ref{app:HSpctrSym}):
\begin{equation}
\mathcal{E}_1 (\Omega,\Delta,B) = - \mathcal{E}_m (\Omega,-\Delta,-B). \label{eq:E1msym}
\end{equation}

%%%%%%%%%%%%%%%%%%%%%%%%%%%%%%%%%%%%%%%%%%%%%%%%
\begin{figure}[t]
\includegraphics[width=8.5cm]{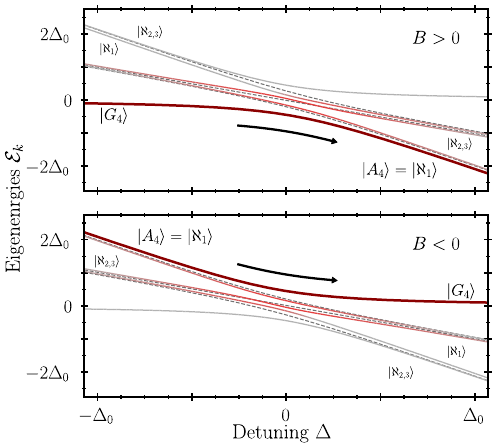}
    \caption{Energy eigenvalues $\mathcal{E}_k$ of the full Hamiltonian $\mathcal{H}$ vs detuning $\Delta$, for $\nu = 4$ atoms interacting via the van der Waals interaction with $B>0$ (top) and $B<0$ (bottom). The parameters are $|B| = 5\Delta_0$ and $\Delta_0 = 3\Omega$.
    The style and colors of the lines have the same meaning as in Fig.~\ref{fig:PXPspectrum}. 
    For odd $\nu = 3,5$, the spectra are similar to those in Fig.~\ref{fig:PXPspectrum}. }
    \label{fig:vdWspectrumN4}
\end{figure}
%%%%%%%%%%%%%%%%%%%%%%%%%%%%%%%%%%%%%%%%%%%%%%%%

The situation is more involved for even $\nu$, as detailed in Appendix~\ref{app:AFMstatevdW} and illustrated in Fig.~\ref{fig:vdWspectrumN4} for $\nu=4$ and relatively large $|B|$ (and $|B_2|$).
As mentioned above, the manifold of AFM-like states with $\nu_r=\nu/2$ Rydberg excitations contains $\nu_r+1$ configurations, but now the long-range interaction of Eq.~(\ref{eq:LRAAHam}) partially lifts the degeneracy of these configurations, since the two 
ordered configurations $\ket{r1r1 \ldots r1}$ and $\ket{1r1r \ldots 1r}$ are shifted by $B_2 (\nu_r - 1)$ while the remaining 
$(\nu-2)/2$ configurations $\ket{r1\ldots r11r \ldots 1r}$ with a defect at various positions in the bulk are shifted by $B_2 (\nu_r - 2)$. 
For positive $B,B_2>0$ and $\Delta \gg  \Omega_0$ the lowest energy AFM-like state $\ket{\alpha_1} = \ket{A_\nu}$ then consists of a proper superposition $\ket{\aleph_1}$ of the defect configurations all having the energy
\[
\mathcal{E}^{(2)}_\mathrm{o} = -\nu_r ( \Delta + S ) + (\nu_r-1) (B_2 -S_{2B}) - S_B ,   
\]
and coupled by the second-order hopping $J=S+S_B$ of Rydberg excitation next to the defect, see Appendix~\ref{app:AFMstatevdW}. 
But for $-\Delta \gg  \Omega_0$ the same state $\ket{A_\nu} \neq \ket{\alpha_m}$ is not the highest energy state $\ket{\alpha_m}$ that is adiabatically connected to $\ket{G_\nu}$ as we sweep $\Delta$ from $-\Delta_0$ to $\Delta_0$, see Fig.~\ref{fig:vdWspectrumN4}.  
Rather, the highest energy states are the symmetric and antisymmetric superpositions $\ket{\aleph_{\nu/2,\nu/2+1}}$ of the two ordered configurations. 
On the other hand, for negative $B,B_2<0$ and $-\Delta \gg  \Omega_0$, the state $\ket{A_\nu} = \ket{\alpha_m}$ is the highest energy AFM-like state adiabatically connected to $\ket{G_\nu}$ as we sweep $\Delta$ from $-\Delta_0$ to $\Delta_0$. 
Moreover, taking into account the second-order level shifts $S_B$ and $S_{2B}$, we again recover the symmetry of the energy eigenvalues $\mathcal{E}_{1}$ and $\mathcal{E}_{m}$ in Eq.~(\ref{eq:E1msym}).

\subsubsection{Dynamics of the system}
\label{sec:vdWdyn}

%%%%%%%%%%%%%%%%%%%%%%%%%%%%%%%%%%%%%%%%%%%%%%%%
\begin{figure}[t]
\includegraphics[width=\columnwidth]{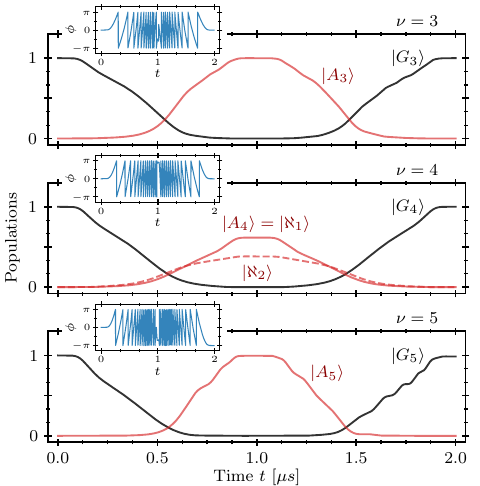}
    \caption{Dynamics of populations of states $\ket{G_\nu}$ and $\ket{A_\nu}$ (solid lines) for the chains of $\nu= 3, 4, 5$ atoms (upper, middle, lower panels) interacting via the van der Waals interaction, as described by the full Hamiltonian $\mathcal{H}$, with the pulse parameters as in Fig.~\ref{fig:PulsesPopulations}(a) and interaction $|B|= 2\pi \times 45\:$MHz: $B>0$ for $0<t\leq \tau =1\:\mu$s, and $B<0$ for $\tau <t\leq 2\tau$. 
    For $\nu=4$ (middle panel) we also show the population of the first excited state $\ket{\aleph_2}$ (dashed line); the sum of populations of states $\ket{\aleph_1} = \ket{A_4}$ and $\ket{\aleph_2}$ is close to 1. 
    Insets show the total phase $\phi(t) = \arg\bra{\Psi(0)}\ket{\Psi(t)}$ acquired by the state of the system during the evolution; The final phase $\phi(2\tau) = 0$ for $\nu=3,4$ ($\nu_r=2$) and $\phi(2\tau) = \pi$ for $\nu=5$ ($\nu_r=3$).}
    \label{fig:PopulationsVdW}
\end{figure}
% %%%%%%%%%%%%%%%%%%%%%%%%%%%%%%%%%%%%%%%%%%%%%%%

We now consider the dynamics of the system described by the full Hamiltonian $\mathcal{H}$. 
The atoms are subject to the same pulses as in Eq.~(\ref{eqs:PulseProfile}) and Fig.~\ref{fig:PulsesPopulations}(a) and we assume that during the second pulse, $\tau < t \leq 2\tau$, the sign of the interatomic interaction $B'=-B$ is opposite to that during the first pulse, $0 < t \leq \tau$. 

One possibility to change the sign of $B$ is to quickly transfer the atoms in the Rydberg state $\ket{r}$ to another Rydberg state $\ket{r'}$ with appropriate van der Waals coefficient $C_6' = -C_6$.
This can be done with optical (Raman) pulses coupling states $\ket{r}=\ket{nS}$ and $\ket{r'}=\ket{n'D}$ via an intermediate lower lying state, e.g., $\ket{5P}$ or $\ket{6P}$ for Rb or Cs, while avoiding strong resonant dipole-dipole interactions between the atoms in the Rydberg states. 
Using properly tuned laser pulses, the transition $\ket{r}\rightarrow\ket{r'}$ can be executed with high fidelity (see below) in $t_{\mathrm{tr}}<0.1\:\mu$s ($t_{\mathrm{tr}} \ll \tau$) under realistic experimental conditions, as discussed in Appendix~\ref{app:BtomB}. 

In Fig.~\ref{fig:PopulationsVdW} we show the dynamics of the system of $N=5$ atoms with the two end atoms representing qubits in state $\ket{0} \ket{0}$ and thus $\nu=3$ (top panel), in state $\ket{0} \ket{1}$ or $\ket{1} \ket{0}$ and thus $\nu=4$ (middle panel), and $\ket{1} \ket{1}$ and thus $\nu=5$ (bottom panel). 
Note that for $\nu=4$, at the end of the first pulse, $\Delta(t=\tau) = \Delta_0$, the population of the lowest energy (defect) state $\ket{\alpha_1} = \ket{A_4} = \ket{\aleph_1}$ with $\nu_r=2$ Rydberg excitations and energy $\mathcal{E}_1 = -2(\Delta + S + S_B)$ is significantly smaller than unity, while the remaining population is mostly accumulated in the other state $\ket{\alpha_2}=\ket{\aleph_2}$ with the same $\nu_r=2$ Rydberg excitations and slightly larger (by about $B_2$) energy $\mathcal{E}_{2} = -2(\Delta + S) + B_2 -S_{2B} - S_B$ (state $\ket{\aleph_3}$ with the same $\mathcal{E}_3= \mathcal{E}_2$ and $\nu_r=2$ is dark and therefore not populated), see Appendix~\ref{app:AFMstatevdW}.      
As we flip the sign of $B \to B'=-B$ and apply identical pulse with $\Delta(t=\tau+0) = -\Delta_0$, the energies of the corresponding eigenstates are reflected about the $\mathcal{E} = 0$ and $\Delta=0$ axes, i.e., $\mathcal{E}_{1,2,\ldots}(\Omega,\Delta_0,B) = -\mathcal{E}_{m,m-1,\ldots}(\Omega,-\Delta_0,-B)$.
The system now undergoes, up to the parity $(-1)^{\nu_r}$ (see Appendix~\ref{app:parityphaseVdW}), the time-reversed dynamics and ends up in state $\ket{G_\nu}$ for $\Delta = \Delta_0$ at $t=2\tau$, while the dynamical phases $\phi_d^{(\mathrm{I,II})}$ during the first (I) and second (II) steps fully cancel each other since 
\[
\phi_\mathrm{d}^{(\mathrm{I})} = \int_{0}^\tau \! \langle\mathcal{H}(t)\rangle dt = - \int_{\tau}^{2\tau} \!\! \langle\mathcal{H}(t)\rangle dt = - \phi_\mathrm{d}^{(\mathrm{II})} .
\]

Note that if the absolute value of the interaction during the first and the second steps are not exactly the same, $B'= -\chi B$ ($C_6'= -\chi C_6$) with $\chi \neq 1$, we can rescale the parameters of the second pulse as $\Omega_0'= \chi \Omega_0$ and $\Delta_0'= \chi \Delta_0$ and its duration as $\tau' = \tau/\chi$ to achieve exactly the same unitary transformation that returns the system to the initial state $\ket{G_\nu}$ and cancels the dynamical phase, provided the atoms do not move much and their relative positions (not necessary fully uniform) can be assumed constant during the total gate time $\tau_{\mathrm{tot}} = \tau+\tau'$. Otherwise, we will encounter phase errors as discussed in Sec.~\ref{subsec:Thermot}.

\section{Gate fidelity}
\label{sec:gateFE}

We now analyze the performance of our scheme which we quantify via the error probability $E=1-F$, where $F$ is the gate fidelity averaged over all the input states of the qubits.
For a two-qubit gate, the average fidelity is \cite{pedersen2007fidelity}
\begin{equation}
F = \frac{1}{20}[\Tr{\mathcal{M}\mathcal{M}^\dagger}+|\Tr{\mathcal{M}}|^2] \label{eq:FidelityDef}
\end{equation}
with $\mathcal{M} = \mathcal{U}_\mathrm{CZ}^{\dagger} \mathcal{U}$,  where
\[
\mathcal{U}_{\text{CZ}} = 
\begin{pmatrix}
    1 & 0& 0& 0\\
    0 & 1& 0& 0\\
    0 & 0& 1& 0\\
    0 & 0& 0& -1
\end{pmatrix} 
\]
is the transformation matrix of the ideal \textsc{cz} gate and $\mathcal{U}$ is the actual transformation of the two qubits resulting from our protocol. 
To determine $\mathcal{U}$, we solve the Schrödinger equation $i\partial_t \ket{\Psi} = \tilde{\mathcal{H}} \ket{\Psi}$ for the $N$-atom system initially in state $\ket{\Psi(0)} = \ket{q_1}\ket{11 \ldots 1} \ket{q_2}$ ($q_{1,2} = 0,1$) using the effective Hamiltonian 
\begin{equation}
    \tilde{\mathcal{H}} = \mathcal{H} - \frac{i}{2}\mathcal{L}^2 \label{eq:HamTotEff}
\end{equation}
containing the Hermitian part $\mathcal{H} = \mathcal{H}_\mathrm{aa} + \mathcal{H}_\mathrm{af}$ and the non-Hermitian part $\mathcal{L}^2 =\sum_{\rho =  r,r'}\sum_{i=1}^N \Gamma_\rho\ket{\rho_i}\bra{\rho_i}$ that describes the decay of the Rydberg states $\ket{\rho}$ ($\rho = r,r'$)  with rates $\Gamma_{\rho}$ and thereby reduces the norm of $\ket{\Psi(t)}$ during the evolution.   
After the evolution, at time $t=\tau_{\mathrm{tot}} = \tau + \tau'$, we project $\ket{\Psi(\tau_{\mathrm{tot}})}$ onto the subspace spanned by the initial states, $P_{q_1} \otimes P_{\mathrm{bus}} \otimes P_{q_2}$, where  
$P_{q_{1,2}} = \ket{0}_{1,2}\bra{0} + \ket{1}_{1,2}\bra{1}$
and $P_{\mathrm{bus}} = \ket{1} \bra{1}^{\otimes (N-2)}$.
Hence, from the four output states corresponding to the four possible input states of the two qubits, we obtain the transformation matrix $\mathcal{U}$ which is diagonal, since $\mathcal{H}$ does not couple the qubit states $\ket{0,1}$, and non-unitary, since we project onto the subspace of the initial states and assume that decay from the Rydberg states takes the atoms outside the computational subspace. 
Neglecting the decay of the Rydberg states back to the initial states and projecting the state vector $\ket{\Psi(\tau_{\mathrm{tot}})}$ onto the computational subspace overestimates the gate error leading to the lower bound for the fidelity, but it simplifies the computation, as we do not need to average over many quantum jump trajectories \cite{PLDP2007,Plenio1998}. 

\subsection{Decay and leakage errors}
\label{subsec:decayLZ}

There are two main sources of errors stemming from the atomic decay and non-adiabatic transitions, or leakage, to the states with the wrong parity.

During time $\tau_{\mathrm{tot}} = \tau +\tau'$, the decay error averaged over all the input states of the two qubits  is 
\begin{equation}\label{eq:Edecay}
E_\mathrm{decay} = 1 - \exp( - \tfrac{1}{2}\bar{\nu}_r \Gamma \tau_{\mathrm{tot}}) \simeq \tfrac{1}{2}  \bar{\nu}_r \Gamma \tau_{\mathrm{tot}} ,
\end{equation}
where $\bar{\nu}_r = N/2 - 1/4$ is the mean number of Rydberg excitations, $\Gamma = (\Gamma_r \tau + \Gamma_{r'} \tau')/\tau_{\mathrm{tot}}$ is the mean decay rate of the atoms in the Rydberg states $\ket{r}$ (in step I) and $\ket{r'}$ (in step II), and the factor $1/2$ signifies the fact that the atoms spend approximately half of the total time $\tau_{\mathrm{tot}}$ in the Rydberg states. 

The non-adiabatic transition, or leakage, to the states with wrong parity during steps I and II results in the reduction of population $|g|^2 \simeq 1-4|b|^2$ of state $\ket{G_\nu}$ at time $\tau_\mathrm{tot}$, where $|b|^2$ is the population transferred to the wrong parity states during step I, see Appendix~\ref{app:Eleakagewp}.
The dominant wrong-parity state is the state $\ket{\alpha_k}$ ($k=2$ for $\nu$ odd, and $k=\nu/2+2$ for $\nu$ even) that tends asymptotically (for $\Delta \gg |\Omega|$) to a configuration with one less Rydberg excitation than the target lowest energy state $\ket{\alpha_1} \to \ket{A_{\nu}}$ with $\nu_r= \lceil \nu/2 \rceil$ Rydberg excitations. 
We can estimate the non-adiabatic transition probability to this state using the Landau-Zener formula \cite{Tzortzakakis2022Microscopic} 
\begin{equation}
    |b|^2 \simeq p_\mathrm{LZ} (\nu) = \exp \left [-2\pi \frac{(\delta \mathcal{E}_\nu/2)^2}{2\Delta_0/\tau}  \right] ,
\end{equation}
where $\delta \mathcal{E}_\nu \equiv \min |\mathcal{E}_1 - \mathcal{E}_k|$ is the minimal energy gap between the states $\ket{\alpha_1}$ and $\ket{\alpha_k}$ 
in the vicinity of $\Delta = 0$. 
The leakage error averaged over the $2^2$ input states of two qubits, $\ket{q_1} \ket{q_2}$ with $q_{1,2} = 0,1$ and therefore $\nu = N-2,N-1,N$, is thus
\begin{equation}\label{eq:Eleakgen}
E_\mathrm{leakage} = \tfrac{4}{2^2} [p_\mathrm{LZ} (N-2) + 2 p_\mathrm{LZ} (N-1) + p_\mathrm{LZ} (N)] .
\end{equation}

The energy gap $\delta \mathcal{E}_\nu \equiv \min |\mathcal{E}_1 - \mathcal{E}_k| = \kappa_\nu \Omega_0$ is smaller for $\nu$ odd ($k=2$) than for $\nu$ even ($k=\nu/2+2$).
Here the dimensionless parameter $\kappa_\nu$ depends only on $\nu$: 
$\kappa_\nu \propto \nu^{-p}$ with $p <1$ for $\nu \lesssim 10$ due to the boundary effects \cite{Tzortzakakis2022Microscopic} and $p=1$ in the thermodynamic limit $\nu \to \infty$ of the Ising model with finite-range interactions.  
Hence, the dominant contribution to the leakage error in Eq.~(\ref{eq:Eleakgen}) comes from the term $p_\mathrm{LZ} (\nu)$ with the largest odd $\nu$ for the given total number of atoms $N$, i.e., $p_\mathrm{LZ} (N)$ for $N$ odd and $p_\mathrm{LZ} (N-1)$ for $N$ even. 
The leakage error can then be approximated as
\begin{equation} \label{eq:Eleak}
E_\mathrm{leakage} \approx \mu p_{\mathrm{LZ} }(\nu) = 
\mu \exp \left( - c_\nu \frac{\Omega_0^2}{\Delta_0} \tau \right) ,
\end{equation}
where $\mu = 1$ and $\nu = N$ for $N$ odd and $\mu = 2$ and $\nu = N-1$ for $N$ even, while $c_\nu = \pi \kappa_{\nu}^2/4$.  
We determine the values of $c_\nu$ via fitting the solution of the Schrödinger equation using the full Hermitian Hamiltonian $\mathcal{H}$ (omitting the decay).
%The values of $c_N$ estimated via fitting without decay ($\Gamma=0$) and $\Delta_0/|B| = 0.444$, $\Omega_0/|B| =0.178$ are:
%$c_3 = 0.431(6)$, $c_4 = 0.432(4)$, $c_5 = 0.280(4)$, $c_6 = 0.283(3)$, $c_7 = 0.190(3)$ and $c_8 = 0.197(3)$. 

%%%%%%%%%%%%%%%%%%%%%%%%%%%%%%%%%%%%%
\begin{figure}[t]
\includegraphics[width=8.7cm]{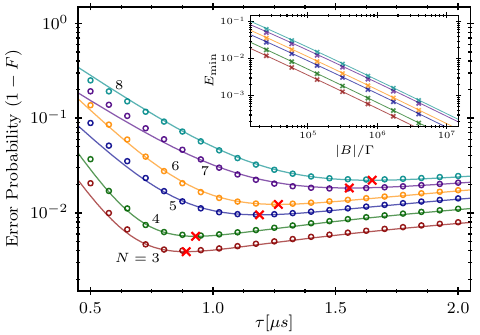}
\caption{Total error probability (infidelity) $E=1-F = E_{\mathrm{decay}} + E_{\mathrm{leakage}}$ vs the pulse duration $\tau$ for chains of $N=3-8$ atoms (from bottom to top) as obtained from Eqs.~(\ref{eq:Edecay}) and (\ref{eq:Eleak}) (solid lines) and exact numerical solution of the    
the Schrödinger equation with Hamiltonian (\ref{eq:HamTotEff}) (open circles).
The optimal $\tau_{\mathrm{opt}}$'s (red crosses) are from Eq.~(\ref{eq:tauopt}). 
The parameters are as in Figs.~\ref{fig:PulsesPopulations} and \ref{fig:PopulationsVdW}: $|B| = 2\pi \times 45\,$MHz, $\Delta_0 = 2\pi \times 20\,$MHz, and $\Omega_0 = 2\pi \times 8\,$MHz, resulting in $c_{3,5,7} \simeq (0.43, 0.28, 0.19)$ in Eq.~(\ref{eq:Eleak}), while $\Gamma_{r,r'} = 2\pi \times 0.5\,$kHz in Eq.~(\ref{eq:Edecay}).
Inset shows the scaling of the minimal error $E_{\mathrm{min}}$ with $|B|/\Gamma$ for $N = 3-8$ (from bottom to top) as obtained from Eq.~(\ref{eq:EminvsB}) (solid lines) and exact numerical simulations (crosses) with the ratios $\lambda_1 = \Omega_0/|B| =8/45$ and $\lambda_2 = \Delta_0/|B| =20/45$ assumed fixed. }
\label{fig:ErrNvstau}
\end{figure}
%%%%%%%%%%%%%%%%%%%%%%%%%%%%%%%%%%%%%

To reduce decay error (\ref{eq:Edecay}), we need pulses of short duration $\tau$, while to suppress non-adiabatic leakage error (\ref{eq:Eleak}), we need longer pulses. 
Our aim is thus to minimize the total error $E=E_\mathrm{decay} + E_\mathrm{leakage}$.
This is achieved for the pulse duration
\begin{equation}
    \tau_\mathrm{opt} \simeq \frac{1}{c_\nu} \frac{\Delta_0}{\Omega_0^2} \, \ln \left(\frac{\mu c_\nu}{\bar{\nu}_r} \frac{\Omega_0^2}{\Gamma \Delta_0}  \right) , 
    \label{eq:tauopt}
\end{equation}
leading to the minimal error
\begin{equation}
E_{\min} \simeq 
\frac{\bar{\nu}_r}{ c_\nu} \frac{\Gamma \Delta_0}{\Omega_0^2} \left[1 + \ln \left(\frac{\mu c_\nu }{\bar{\nu}_r} \frac{\Omega_0^2}{\Gamma \Delta_0}  \right) \right]  ,
\end{equation}
and we assume $\tau_{\mathrm{tot}} \simeq 2 \tau$. 
We have verified these conclusions via exact numerical simulations for $N \leq 8$ atoms, see Fig.~\ref{fig:ErrNvstau}.

Taking into account the conditions $\Omega_0 < \Delta_0 < B$, we can express the laser parameters through the van der Waals interaction as $\Omega_0 = \lambda_1 B$ and $\Delta_0 = \lambda_2 B$, where $\lambda_1 < \lambda_2 < 1$, obtaining
for the minimal error 
\begin{equation}
E_{\min} \simeq \frac{\bar{\nu}_r \lambda_2}{ c_\nu\lambda_1^2} \frac{\Gamma}{B} \left[1 + \ln \left(\frac{\mu c_\nu \lambda_1^2}{\bar{\nu}_r\lambda_2} \frac{B}{\Gamma}  \right) \right] , \label{eq:EminvsB}
\end{equation}
which is shown in the Inset of Fig.~\ref{fig:ErrNvstau}
assuming fixed values of $\lambda_{1,2}$.

For larger $N \gg 1$, the minimal energy gap $\delta \mathcal{E}_\nu \propto \Omega_0/\nu$ is approximately the same for $\nu = N, N-1, N-2$ and all the terms in Eq.~(\ref{eq:Eleakgen}) contribute nearly equally to the non-adiabatic leakage error in Eq.~(\ref{eq:Eleak}) with $\mu \simeq 4$.
With $\bar{\nu}_r \simeq N/2$ and $c_\nu \simeq c_N  \simeq c/N^2$ we then have 
\[
E_{\min} \simeq C N^3 \frac{\Gamma}{B} \left[ 1 + \ln \left(\frac{4}{C N^3} \frac{B}{\Gamma} \right) \right] ,
\]
where $C = \lambda_2 / (2c\lambda_1^2)= \mathcal{O}(1)$. 
For a pair of qubits separated by distance $L$, the lattice constant is $a\simeq L/N$ and we can express the nearest-neighbor van der Waals interaction as $B\simeq C_6 N^6/L^6$, which, upon substitution in the above equation, leads to the error scaling $E_{\min} \propto \Gamma L^6/(C_6 N^3)$, where we have omitted the slowly-varying logarithmic dependence. Hence, for a given distance $L$, increasing the number of atoms $N$ in the quantum bus will improve the gate fidelity.
We finally note that for the dipole-dipole interaction $B\simeq C_3 N^3/L^3$, the same arguments would lead to $E_{\min} \propto \Gamma L^3/C_3$.

\subsection{Rydberg state transfer error}
\label{subsec:rrptransfer}

So far we have neglected errors due to transfer between the Rydberg states $\ket{r}$ and $\ket{r'}$ with opposite signs of interactions $B \simeq -B'$ between steps I and II. 
As already mentioned in Sec.~\ref{sec:vdWdyn}, this transfer can be implemented very fast, $t_{\mathrm{tr}} \ll \tau$, and therefore the probability of Rydberg state decay during the transfer, $\sim \bar{\nu}_r \Gamma t_{\mathrm{tr}}$, is negligible compared to $E_{\mathrm{decay}}$ in Eq. (\ref{eq:Edecay}).
Of course the transfer itself can lead to error, which should be much smaller than $E_{\min}$ of Eq.~(\ref{eq:EminvsB}).

In Appendix~\ref{app:BtomB} we argue that an experimentally realistic way to perform the transfer with little error is to use optical (Raman) pulses to couple the Rydberg states $\ket{r} = \ket{n_S S_{1/2}}$ and $\ket{r'} = \ket{n_D D_{5/2}}$ via an intermediate non-resonant lower-lying state $\ket{5P_{3/2}}$ or $\ket{6P_{3/2}}$ of the Rb or Cs atom. 
For the effective (two-photon) Rabi frequency $\Omega_{SD}$ and detuning $\delta_{SD}$ of the lasers coupling states $\ket{r}$ and $\ket{r'}$, we find that transfer error per atom initially in the Rydberg state $\ket{r}$ is either $E_{\mathrm{bulk}} \simeq |\delta_{SD} + (B_2-B_2')|^2/|\Omega_{SD}|^2$ if the atom is in the AFM bulk (has two next-nearest neighbors), or $E_{\mathrm{edge}} \simeq |\delta_{SD} + (B_2-B_2')/2|^2/|\Omega_{SD}|^2$ if the atom is at the edge or near the AFM defect (has only one next-nearest neighbor), and $B_2=B/64$ and $B_2'=B'/64$ are the next-nearest neighbor interactions. 
Setting $\delta_{SD} = -(B_2-B_2')$ or $\delta_{SD} = -(B_2-B_2')/2$ we have either $E_{\mathrm{bulk}} \simeq 0$ and $E_{\mathrm{edge}} \simeq |B_2/\Omega_{SD}|^2 \ll 1$ or the other way around, $E_{\mathrm{edge}} \simeq 0$ and $E_{\mathrm{bulk}} \simeq |B_2/\Omega_{SD}|^2 \ll 1$, assuming $B_2 \simeq -B_2' \ll \Omega_{SD}$. 
With experimentally realistic parameters, the largest transfer error per atom can be as small as $E_{\mathrm{edge,bulk}} \simeq |B_2/\Omega_{SD}|^2 < 10^{-2}$.
  
Hence, for a configuration with $\nu_{\mathrm{edge}}$ edge atoms and $\nu_{\mathrm{bulk}}$ atoms, the total transfer error is $E_{\mathrm{transfer}} \simeq \nu_{\mathrm{edge}} E_{\mathrm{edge}} + \nu_{\mathrm{bulk}} E_{\mathrm{bulk}}$. For long chains with many atoms in the bulk, $\nu_{\mathrm{bulk}} > \nu_{\mathrm{edge}} = 2(4)$, we can then set $\delta_{SD} \simeq -2B_2$, while for shorter chains we can chose the detuning $-2B_2 < \delta_{SD} < B_2$ to minimize the total transfer error $E_{\mathrm{transfer}} \lesssim 0.01$, which still remains small compared to $E_{\min}$ of Eq.~(\ref{eq:EminvsB}).

\subsection{Thermal motion}
\label{subsec:Thermot}

As mentioned above, small static disorder in the atomic positions does not affect the gate fidelity since by changing the sign of the van der Waals interaction coefficient $C_6' = - C_6$ in step II, we reverse the signs of all the interactions, $B_{ij}' = - B_{ij} \; \forall \; i,j$, and thereby the dynamics of the system, which, due to the symmetry of the spectrum (see Appendix~\ref{app:HSpctrSym}) returns to the initial state $\ket{G_{\nu}}$ and acquires the appropriate geometric phase while the dynamical phase is fully canceled (see Appendix~\ref{app:parityphaseVdW}). 

But if the atoms move during and between steps I and II, the interactions will not satisfy the condition $B_{ij}' = - B_{ij}$ (or $B_{ij}' = - \chi B_{ij}$ with constant $\chi$) and the dynamical phases will not completely cancel, resulting in dephasing. 
For atoms at temperature $T$, the most probable thermal velocity along their separation direction is $v_{\mathrm{th}} = \sqrt{k_\mathrm{B} T/m_{\mathrm{Rb}}}$, where $k_\mathrm{B}$ is the  Boltzmann constant and $m_{\mathrm{Rb}}$ is the atomic mass, assumed that of $^{87}$Rb. 
Since in the AFM-like state of Rydberg excitations the dominant interatomic interaction is the next-nearest neighbor interaction $B_2 = C_6/(2a)^6$, we require that its change $\delta B_2 \simeq B_2 \, 3\sqrt{2}v_{\mathrm{th}} \tau/a$ during time $\tau$ be sufficiently small, so that the resulting interaction-induced phase fluctuations for a pair Rydberg atoms $\delta \phi = \frac{1}{2}\delta B_2 \tau \ll \pi$.  
For $\nu_r \gg 1$ atoms in the Rydberg state, the phase fluctuations in the bulk of AFM-like state of Rydberg excitations are averaged out to first order in $v_{\mathrm{th}} \tau/a$, and only the two edge atoms, and possibly the pair of atoms next to the defect in the AFM-like bulk for even $N$, contribute to the dephasing $\delta \phi$, which we verified via classical Monte-Carlo simulations for atomic position variations due to thermal motion.
With $B_2 = B/64 \lesssim 2\pi \times 1\;$MHz and $a \simeq 4\:\mu$m, we then obtain $\delta \phi \approx 2B_2v_{\mathrm{th}} \tau^2/a \lesssim 0.03\:$rad for $\tau=1\:\mu$s and $T \lesssim 1\:\mu$K which is typical in most experiments \cite{evered2023high,graham2019,graham2022,Levine2019Parallel}. The resulting loss in fidelity of Eq.~(\ref{eq:FidelityDef}) is about $2\times 10^{-4}$.
Decreasing the pulse duration $\tau$ will further decrease the dephasing error due to the thermal motion, as the corresponding error scales as $E_T \propto \delta \phi^2 \propto \tau^4$.

\subsection{Fast and robust gate with optimized pulses}
\label{subsec:Optmization}

We can further improve the gate performance by using quantum optimal control to determine the temporal profile of the laser pulse driving the atoms. 
Following \cite{jandura2022time}, we use the Gradient Ascent Pulse Engineering (GRAPE) method \cite{Khaneja2005GRAPE}, in combination with the Broyden–Fletcher–Goldfarb–Shanno (BFGS) algorithm for unconstrained nonlinear optimization \cite{nocedal1999numerical, Virtanen2020SciPy}, to find the pulses that maximize the gate fidelity.

Since the decay error is proportional to the time the system spends in the Rydberg state, we should find the pulses that can transfer the atoms to the Rydberg state and back in shortest possible times. 
The pulse must have the largest possible area $\theta = \int_0^\tau \Omega(t)dt$ with $\max (\Omega)$ limited by the experimental considerations. 
On the other hand, the best pulse is the shortest pulse, with smallest pulse area $\theta$, that still leads to gate (parity) errors smaller that the decay errors.  
We thus take the same flat-top temporal profile of $\Omega(t)$ as before and optimize the pulse detuning $\Delta(t)$, or its phase $\varphi(t) = \int_0^t \Delta(t') dt'$, while imposing the symmetry conditions $\Delta(t+\tau) = - \Delta(\tau-t)$ (for $0<t \leq \tau$) and $B'=-B$ between the second and first steps to cancel the dynamical phases. 
Assuming small decay rate $\Gamma \tau \ll 1$, the non-Hermitian part of Hamiltonian (\ref{eq:HamTotEff}) does not significantly affect the time evolution of the system. 
We can therefore optimize the gate with $\Gamma = 0$ and then include the decay to determine the total fidelity. 

%%%%%%%%%%%%%%%%%%%%%%%%%%%%%%%%%%%
\begin{figure}
    \centering
    \includegraphics[width=\linewidth]{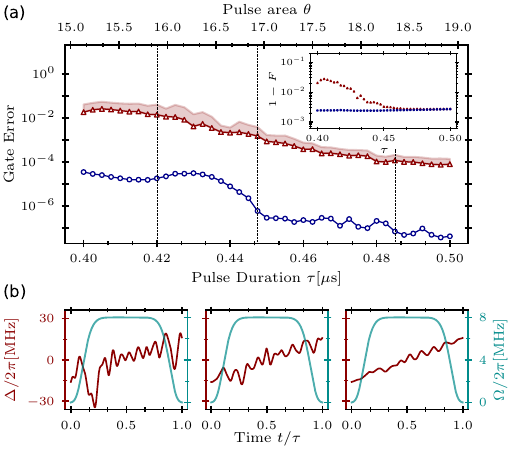}
    \caption{(a) Gate error with optimal pulses vs pulse duration $\tau$ (lower horizontal axis) or pulse area $\theta = \int_0^\tau \Omega(t) dt$ (upper horizontal axis), for $N=5$ atoms and other parameters as in Figs.~\ref{fig:PulsesPopulations}, \ref{fig:PopulationsVdW}, \ref{fig:ErrNvstau}.
    Main panel shows the gate (leakage or parity) error in the absence of atomic decay, $\Gamma_{r,r'} =0$, and position uncertainly  (blue open circles), and with (Gaussian) position uncertainty  $\sigma_{x,y,z} = 0.01a$ (standard deviation), corresponding to random variation $\delta B \lesssim  0.06B$ of the interatomic interaction, as obtained from 150 independent realizations (red open triangles, with shading corresponding to one standard deviation). 
    Inset shows the total error $(1-F)$ including the Rydberg state decay $\Gamma_{r,r'} = 2\pi \times 0.5\:$kHz as in Fig.~\ref{fig:ErrNvstau}, without (blue filled circles) and with (red filled triangles) atomic position uncertainty.   
    (b) Time dependence of the pulse amplitude $\Omega(t)$ and optimized detuning $\Delta(t)$ for three different pulse durations $\tau=0.42,0.447,0.485\:\mu$s indicated by the vertical dashed lines in (a).}
    \label{fig:OptResults}
\end{figure}
%%%%%%%%%%%%%%%%%%%%%%%%%%%%%%

Considering again the chain of $N=5$ atoms, we start with a pulse of large area $\theta_\mathrm{max} \simeq 19$ and linear sweep of the detuning as initial guess. 
We apply the BFGS method to modify $\Delta(t)$ until the change in fidelity $F$ between consecutive iterations is $\delta F < 10^{-12}$ which signifies convergence. 
We then decrease the pulse area by $ \delta \theta \simeq 0.1$ and repeat the procedure using as initial guess the optimal pulse of the previous step, as so on till $\theta_\mathrm{min} \simeq 15$. 
In Fig.~\ref{fig:OptResults} we show the results of optimization that minimize the gate error assuming perfectly ordered lattice. 
For sufficiently long pulses $\tau \gtrsim 0.45\:\mu$s corresponding to pulse area $\theta \gtrsim 17$, we achieve gate error $1-F \lesssim 10^{-7}$ limited by the accuracy of integration method for solving the Schrödinger equation. 
For smaller values of $\theta<17$, or shorter pulses with the same $\max (\Omega)$, the gate error rapidly increases. 
Hence, such short pulses cannot accomplish high fidelity transfer of the atoms to the target AFM-like state of Rydberg excitations with party $\nu_r$ and back even upon optimization.

Since the found optimal pulses are not adiabatic, the transfer to the target state may involve resonances with the excited many-body states of the system.  
The gate fidelity obtained with such pulses is therefore sensitive to the precise value of the nearest-neighbor interaction $B$ and thus to the atomic positions.
Small variations of the atomic positions, leading to variation $\delta B$ in the interatomic interaction strengths, indeed significantly reduce the gate fidelity, 
see Fig.~\ref{fig:OptResults}(a). 
But even with the experimentally relevant atomic position uncertainty $\sigma_{x,y,z}= 0.01 a$, corresponding to $\delta B \lesssim  0.06B$, the gate fidelity with the same optimized pulses is still quite high, $1-F \lesssim 3\times10^{-3}$ for $\theta \gtrsim 17$. 

Adding now the decay of the Rydberg state, $\Gamma = 2\pi \times 0.5\:\text{kHz}$, we obtain the total infidelity which is only slightly reduced since the pulse duration $\tau <0.5 \:\mu$s is very short, see the Inset in Fig.~\ref{fig:OptResults}(a). 
The resulting infidelity is about three times smaller than that obtained with slow (adiabatic) pulses with duration $\tau \gtrsim 1.4\:\mu$s optimized for leakage and decay errors.
We finally note that faster gates with optimized pulses are more tolerant to thermal atomic motion since during the shorter gate times the atoms change less their positions and thereby interatomic interaction strengths.

\section{Conclusions}
\label{sec:conclud}

To summarize, we have presented a protocol to realize quantum gates between distant atomic qubits in an array of atoms in microtraps. 
Our protocol relies on the Rydberg blockade mechanism that is used in laser-driven atomic quantum simulators to prepare spatially-ordered states of Rydberg excitations of atoms in lattices.
In our protocol, the pair of atoms encoding qubits and the intermediate atoms playing the role of a quantum bus connecting the qubits are transferred by global frequency-chirped laser pulses to an antiferromagnetic-like state of Rydberg excitations and back to the initial state resulting in state-dependent geometric phase equivalent to the \textsc{cz} gate, while the dynamical phase is canceled for time-symmetric laser pulses and sufficiently cold atoms. 

We performed detailed numerical and analytical analysis of moderately sized systems of $N$ atoms driven by chirped laser pulses.
We determined the spectrum and non-adiabatic dynamics of the system, derived analytical expressions for gate performance and obtained the scaling of gate fidelity with the system size $L$ and atom number $N$. As a byproduct of our study, we acquired detailed understanding of the microscopic structure and energy spectrum of the chain of atoms realizing the quantum Ising model involving finite-strength and long-range interactions.    

With linear frequency chirp and optimal duration of the laser pulses, $2\tau \simeq 2-4\:\mu$s, our scheme can yield gate fidelities $F = 0.995 - 0.98$ for $N=3-8$ atoms, corresponding to interqubit separations of $L \simeq 10-30\:\mu$m in practical experimental setups with atoms in arrays of microtraps separated by $a=3-5\:\mu$s. 
In comparison, moving the atoms trapped in optical tweezers with the slow velocities $v \lesssim 0.5\:\mu\mathrm{m}/\mu\mathrm{s}$ to avoid their loss or decoherence, as is done in Ref. \cite{bluvstein2022quantum}, would require exceedingly long times $2L/v \sim 40-120\:\mu$s to perform quantum gates between the atoms separated by the same distances $L$.   
We have also used quantum optimal control to design phase-modulated pulsed for faster gates with even higher fidelities. 

%%%%%%%%%%%%%%%%%%%%%%%%%%%
\acknowledgments
This work was supported by the EU HORIZON-RIA Project EuRyQa (grant No. 101070144). 

%%%%%%%%%%%%%%%%%%%%%%%%%%%
\appendix

\section{Symmetry of the spectrum}
\label{app:HSpctrSym}

The spectrum of the $PXP$ Hamiltonian, $\mathcal{H}_{PXP} \ket{\alpha_k} = \mathcal{E}_k \ket{\alpha_k}$ ($k=1,2, \ldots , m$), is symmetric with respect to the point $(\mathcal{E}, \Delta) = (0, 0)$, i.e., for any eigenstate $\ket{\alpha_k}$ with eigenenergy $\mathcal{E}_k(\Omega,\Delta)$ there is an eigenstate $\ket{\alpha_{m-k+1}}$ with eigenenergy $\mathcal{E}_{m-k+1}(\Omega,-\Delta) = - \mathcal{E}_k(\Omega,\Delta)$. 

To see this, we define the parity operator 
\begin{equation}
    \mathcal{P} = \prod_{i=1}^\nu ( \ket{1_i}\bra{1_i} - \ket{r_i}\bra{r_i}) ,
\end{equation}
which anticommutes with the first term ($\propto \Omega$) of $\mathcal{H}_{PXP}$ in Eq.~(\ref{eqs:PXPhamiltonian}) and commutes with the second term ($\propto \Delta$). 
This implies that 
$\mathcal{P} \mathcal{H}_{PXP} [\Omega, \Delta] =  \mathcal{H}_{PXP} [-\Omega, \Delta] \mathcal{P}$ or equivalently 
\begin{equation}
    \mathcal{P} \mathcal{H}_{PXP} [\Omega, \Delta] = 
    - \mathcal{H}_{PXP} [\Omega, -\Delta] \mathcal{P}. \label{eq:PHmHPpxp}
\end{equation}
We can then write $\mathcal{P} \mathcal{H}_{PXP} [\Omega, \Delta] \ket{\alpha_k} = \mathcal{E}_k [\Omega, \Delta] \mathcal{P} \ket{\alpha_k} = - \mathcal{H}_{PXP} [\Omega, -\Delta]\mathcal{P}  \ket{\alpha_k}$ 
and therefore 
\[
\mathcal{H}_{PXP} [\Omega, -\Delta]  \ket{\alpha_k'} = \mathcal{E}_k' [\Omega, - \Delta] \ket{\alpha_k'} = - \mathcal{E}_k [\Omega, \Delta] \ket{\alpha_k'} ,
\]
where $\ket{\alpha_k'} = \mathcal{P} \ket{\alpha_k}$ is the eigenstate of $\mathcal{H}_{PXP} [\Omega, -\Delta]$ with eigenvalue $\mathcal{E}_k' [\Omega, - \Delta]$.
Hence, for every eigenstate $\ket{\alpha_k}$ of $\mathcal{H}_{PXP} [\Omega, \Delta]$ with eigenvalue $\mathcal{E}_k [\Omega, \Delta]$, there is indeed an eigenstate $\ket{\alpha_k'}$ of $\mathcal{H}_{PXP} [\Omega, -\Delta]$ with eigenvalue $\mathcal{E}_k' [\Omega, -\Delta] = - \mathcal{E}_k [\Omega, \Delta]$.

For the full Hamiltonian $\mathcal{H} = \mathcal{H}_{\mathrm{af}} + \mathcal{H}_{\mathrm{aa}}$ including the interatomic interactions $\mathcal{H}_{\mathrm{aa}}$ of Eq.~(\ref{eq:AAHam}), the above symmetry is broken, but we can recover it by changing simultaneously the sign of $\Delta$ and $B$ (i.e. $C_6$). 
Indeed, since the party operator commutes with $\mathcal{H}_{\mathrm{aa}}$, we have 
\begin{equation}
    \mathcal{P} \mathcal{H} [\Omega, \Delta,B] = 
    - \mathcal{H} [\Omega, -\Delta,-B] \mathcal{P}. \label{eq:PHmHP}
\end{equation}
Repeating the above steps, we find that for every eigenstate $\ket{\alpha_k}$ of $\mathcal{H} [\Omega, \Delta, B]$ with eigenvalue $\mathcal{E}_k [\Omega, \Delta, B]$, there is an eigenstate $\ket{\alpha_k'}$ of $\mathcal{H} [\Omega, -\Delta, -B]$ with eigenvalue $\mathcal{E}_k' [\Omega, -\Delta,-B] = - \mathcal{E}_k [\Omega, \Delta,B]$.

\section{Antiferromagnetic ground state}
\label{app:AFMstate}

\subsection{\textit{PXP} model}
\label{app:AFMstatePXP}

As stated in Sec.~\ref{subsec:PXP} of the main text, for $\Delta \gg |\Omega|$, the lowest energy eigenstate of the $PXP$ model is the AFM state $\ket{A_\nu}$ with $\nu_r = \lceil \nu/2 \rceil$ Rydberg excitations.  

For $\nu$ odd, in the limit of $\Omega \to 0$, we have a single ordered configuration $\ket{A_\nu} = \ket{r1r1\ldots 1r}$ with $\nu_r = (\nu+1)/2$ excitations and energy $\mathcal{E}^{(0)} = -\nu_r \Delta$. 
A small but finite $\Omega$ couples this configuration to $\nu_r$ configurations $\ket{11r1\ldots 1r}, \ket{r111\ldots 1r},\ldots, \ket{r1\ldots 1r11}$ with one less excitation, all detuned by $\Delta$. 
Since $\Delta \gg |\Omega|$, each of these couplings induces second-order level shift $-|\Omega|^2/(4\Delta) \equiv -S$. 
Then, for the energy of the AFM state we obtain 
\begin{equation}
    \mathcal{E}_1 = -\nu_r ( \Delta + S ) . \label{eq:app:E1PXP}
\end{equation}

For $\nu$ even, in the limit of $\Omega \to 0$, we have $\nu/2+1$ degenerate configurations $\ket{a_1} = \ket{1r1r1 \ldots 1r},
\ket{a_2} = \ket{r11r \ldots 1r}, \ldots , \ket{a_{\nu/2}} = \ket{r1 \ldots r11r}, \ket{a_{\nu/2+1}} = \ket{r1 \ldots r1r1}$ with $\nu_r = \nu/2$ excitations and energy $\mathcal{E}^{(0)} = -\nu_r \Delta$.
The first and the last of these configurations are the ordered AFM configurations, while the remaining ones have a single defect at $(\nu-2)/2$ different positions in the bulk. 
A small but finite $\Omega$ couples each of these configurations to $\nu_r$ configurations with one less excitation, all detuned by $\Delta$. Since $\Delta \gg |\Omega|$, these transitions are suppressed, but together with the level shifts we also obtain the second-order coupling $S$ between the configurations $\ket{a_j}$ and $\ket{a_{j+1}}$ with defect shifted by two lattice sites. 
We can then write an effective Hamiltonian for this low-energy AFM subspace as
\begin{eqnarray}
    \mathcal{H}_{\mathrm{AFM}} &=& - \nu_r(\Delta +S) \sum_{j=1}^{\nu/2+1} \ket{a_j}\bra{a_j} 
    \nonumber \\ & &
    - S \sum_{j=1}^{\nu/2} (\ket{a_j}\bra{a_{j+1}} + \mathrm{H.c.}) . 
\end{eqnarray}
which describes hopping of a defect in an AFM background. 

We can diagonalize the above Hamiltonian, obtaining its eigenstates and the corresponding energy eigenvalues
\begin{subequations}
    \begin{align}
    \ket{\aleph_k}     &= \sqrt{\frac{2}{\nu/2 + 2}} \sum_{j=1}^{\nu/2+1} \sin({\frac{kj\pi}{\nu/2+2}}) \ket{a_j}, \label{subeq:PXPeigenst} \\
   \mathcal{E}_k &= - \nu_r(\Delta +S) - 2S \cos \left( {\frac{k\pi}{\nu/2 + 2}} \right) \label{subeq:PXPeigenval}, 
   \end{align}
\end{subequations}
where $k =1,...,\nu/2+1$. The eigenstate with the lowest energy $\mathcal{E}_1 \simeq - \nu_r(\Delta +S) - 2S$ is $\ket{A_\nu} = \ket{\aleph_1}$ which is populated under the adiabatic conditions. 

%%%%%%%%%%%%%%%%%%%%%%%%%%%%%%%%%%%%%%%%%%%%%%%%
\begin{figure}[t]
    \centering
\includegraphics[width=\columnwidth]{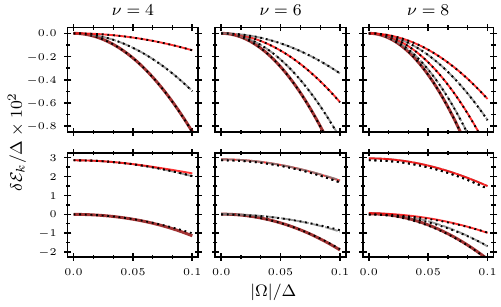}
\caption{Spectrum of the AFM ground-state manifold for $\nu=4,6,8$ atoms and $\Delta \gg |\Omega|$.  
Upper-row panels show $\delta \mathcal{E}_k = \mathcal{E}_k - \mathcal{E}^{(0)}$ for the $PXP$ model as obtained from exact diagonalization of $\mathcal{H}_{PXP}$ (solid and long-dashed lines) and Eq.~(\ref{subeq:PXPeigenval}) (dotted lines). 
Lower row panels show $\delta \mathcal{E}_k = \mathcal{E}_k - \mathcal{E}^{(0)}_\mathrm{d}$ for the full model with $B=3\Delta$ ($B_2= B/64 > S = |\Omega|^2/ (4\Delta)$) as obtained from exact diagonalization of $\mathcal{H}$ (solid and long-dashed lines) and Eq.~(\ref{subeq:vdWeigenvalded}) (dotted lines).}
\label{fig:AFMEexpert}
\end{figure}
% %%%%%%%%%%%%%%%%%%%%%%%%%%%%%%%%%%%%%%%%%%%%%%%

To verify the the above analysis, in Fig.~\ref{fig:AFMEexpert} (upper row) we compare the spectra of the AFM ground-state manifold for different (even) number of atoms $\nu$ as obtained analytically and via exact diagonalization of Hamiltonian $\mathcal{H}_{PXP}$ in Eq.~(\ref{eqs:PXPhamiltonian}). 

As an example that is also used in the main text, consider the chain of $\nu=4$ atoms with $\nu_r=2$ Rydberg excitations. 
In the basis $\{\ket{1r1r}, \ket{r11r}, \ket{r1r1}\}$, the effective Hamiltonian is
\begin{equation}
    \mathcal{H}_{\mathrm{AFM}} = -2 (\Delta +S)  \mathbb{1} 
    -S 
    \begin{pmatrix}
        0 & 1 & 0\\
        1 & 0 & 1\\
        0 & 1 & 0
    \end{pmatrix} 
\end{equation}
having the eigenstates 
\begin{subequations}
    \begin{align}
        \ket{\aleph_1} &= \frac{1}{\sqrt{2}}\ket{r11r} + \frac{1}{2}(\ket{1r1r} + \ket{r1r1}), \label{eq:PXPaleph1}\\
        \ket{\aleph_2} &= \frac{1}{\sqrt{2}}(\ket{1r1r} - \ket{r1r1})\\
        \ket{\aleph_3} &= \frac{1}{\sqrt{2}}\ket{r11r} - \frac{1}{2}(\ket{1r1r} + \ket{r1r1}) 
    \end{align}  
\end{subequations}
and the corresponding energy eigenvalues
\begin{subequations}
 \begin{align}
    \mathcal{E}_1 &= - 2(\Delta+S) - \sqrt{2} S , \\
    \mathcal{E}_2 &= - 2(\Delta+S) , \\
    \mathcal{E}_3 &= - 2(\Delta+S) + \sqrt{2} S .
    \end{align}
\end{subequations}
The antisymmetric state $\ket{\aleph_2}$ is dark as it is decoupled from the laser, while out of the two bright states, the lower energy state is $\ket{A_4} = \ket{\aleph_1}$.

\subsection{van der Waals interacting model}
\label{app:AFMstatevdW}

The above analysis can be generalized for the full model of vdW interacting atoms.
Since the $PXP$ model captures the main physics of the system, for $\Delta \gg |\Omega|$, the lowest energy eigenstate is still the AFM-like state $\ket{A_\nu}$ with $\nu_r = \lceil \nu/2 \rceil$ Rydberg excitations. 
But as stated in Sec.~\ref{subsec:vdW} of the main text, the long range and finite strength of the interatomic vdW interactions $B_{ij}$ leads to the corrections to the $PXP$ model given by Eqs.~(\ref{eq:LRAAHam}), (\ref{eq:HSshift}) and (\ref{eq:HShop}), which have important implications.

We begin again with odd $\nu$. In the limit of $\Omega \to 0$, the single ordered configuration $\ket{A_\nu} = \ket{r1r1\ldots 1r}$ with $\nu_r = (\nu+1)/2$ excitations now has the energy $\mathcal{E}^{(0)} = -\nu_r \Delta + (\nu_r-1) B_2$, the last term coming from the dominant long-range (next nearest-neighbor) interaction of Eq.~(\ref{eq:LRAAHam}).
A small but finite $|\Omega| \ll \Delta, B$ coupling $\ket{A_\nu}$ to states with one less or one more Rydberg excitations induces the second order level shifts as per Eqs.~(\ref{eq:app:E1PXP}) and (\ref{eq:HSshift}), leading to 
\begin{equation}
\mathcal{E}_1 = -\nu_r ( \Delta + S ) + (\nu_r-1) (B_2 -S_{2B}) .  
\end{equation}

For even $\nu$, in the limit of $\Omega \to 0$, we again have $\nu/2+1$ configurations $\ket{a_1},\ket{a_2}, \ldots, \ket{a_{\nu/2+1}}$ with $\nu_r = \nu/2$ excitations. 
But now the long-range interaction of Eq.~(\ref{eq:LRAAHam}) partially lifts the degeneracy: 
The ordered AFM configurations $\ket{a_1}= \ket{1r1r1 \ldots 1r}$ and $\ket{a_{\nu/2+1}} = \ket{r1 \ldots r1r1}$ have energy $\mathcal{E}^{(0)}_\mathrm{o} = -\nu_r \Delta + (\nu_r-1) B_2$, while the energy of $\nu/2-1$ configurations $\ket{a_2} = \ket{r11r \ldots 1r}, \ldots , \ket{a_{\nu/2}} = \ket{r1 \ldots r11r}$ with a defect in the bulk, corresponding to a pair of Rydberg excitations separated by two lattice sites, is lower by $B_2$, $\mathcal{E}^{(0)}_\mathrm{d} = -\nu_r \Delta + (\nu_r-2) B_2$, and we omit the contributions of longer-range interactions $B_{ij}$ with $|i-j| >2$. 

Turning on the laser with $|\Omega| \ll \Delta, B$ that couples $\ket{a_i}$ to the largely detuned states with one less or one more Rydberg excitation, we obtain the second-order level shifts as per Eqs.~(\ref{eq:app:E1PXP}) and (\ref{eq:HSshift}). 
For the two ordered AFM configurations $\ket{a_1},\ket{a_{\nu/2+1}}$ we then have
\begin{equation}
\mathcal{E}^{(2)}_\mathrm{o} = -\nu_r ( \Delta + S ) + (\nu_r-1) (B_2 -S_{2B}) - S_B ,   
\end{equation}
while for the AFM configurations with defect $\ket{a_2}, \ldots , \ket{a_{\nu/2}}$ we have
\begin{equation}
\mathcal{E}^{(2)}_\mathrm{d} = -\nu_r ( \Delta + S ) + (\nu_r-2) (B_2 -S_{2B}) - 2S_B .   
\end{equation}
The second-order hopping of the Rydberg excitation $J=S+S_B$ has now two contributions, the former $S$ due to the virtual de-excitation as before, and the latter $S_B$ due to the virtual excitation of incompletely blockaded atoms, as per Eq.~(\ref{eq:HShop}). 
The effective Hamiltonian for the low-energy subspace of these AFM  configurations is 
\begin{eqnarray}
    \mathcal{H}_{\mathrm{AFM}} &=& 
    \mathcal{E}^{(2)}_\mathrm{o} \!\!\!\!\!\! \sum_{j=1,\nu/2+1} \!\!\!\!\! \ket{a_j}\bra{a_j} +
    \mathcal{E}^{(2)}_\mathrm{d} \sum_{j=2}^{\nu/2} \ket{a_j}\bra{a_j} 
    \nonumber \\ & &
    - J \sum_{j=1}^{\nu/2} (\ket{a_j}\bra{a_{j+1}} + \mathrm{H.c.}) . 
\end{eqnarray}
If $J \gtrsim \mathcal{E}^{(2)}_\mathrm{o} - \mathcal{E}^{(2)}_\mathrm{d}$ ($S \gtrsim B_2$), we can neglects the energy difference between the ordered and defect configurations, $\mathcal{E}^{(2)}_\mathrm{o} \approx \mathcal{E}^{(2)}_\mathrm{d}$, and diagonalize the above Hamiltonian, obtaining the eigenstates and the corresponding energy eigenvalues as 
\begin{subequations}
\label{subeq:vdWeigenpr}
  \begin{align}
  \ket{\aleph_k} &= \sqrt{\frac{2}{\nu/2+2}} \sum_{j=1}^{\nu/2+1} \sin({\frac{kj\pi}{\nu/2+2}}) \ket{a_j}, \\
  \mathcal{E}_k &= \mathcal{E}^{(2)}_\mathrm{d} - 2J \cos \left( {\frac{k\pi}{\nu/2+2}} \right) , \label{subeq:vdWeigenval}
  \end{align}
\end{subequations}
for $k=1,2,\ldots, \nu/2+1$. 
The eigenstate with the lowest energy $\mathcal{E}_1 \simeq \mathcal{E}^{(2)}_\mathrm{d} - 2J$ is $\ket{A_\nu} = \ket{\aleph_1}$, exactly as in Eq.~(\ref{subeq:PXPeigenst}) for the $PXP$ model.

In the opposite limit of $J \ll \mathcal{E}^{(2)}_\mathrm{o} - \mathcal{E}^{(2)}_\mathrm{d}$ ($S \ll B_2$), the ordered AFM configurations $\ket{a_1}$ and $\ket{a_{\nu/2+1}}$ can be assumed decoupled from the defect configurations $\ket{a_2}, \ldots , \ket{a_{\nu/2}}$ due to the large energy mismatch.
Then the eigenstates and the corresponding energy eigenvalues of $\mathcal{H}_{\mathrm{AFM}}$ are
\begin{subequations}
  \begin{align}
  \ket{\aleph_k} &= \sqrt{\frac{2}{\nu/2}} \sum_{j=2}^{\nu/2} \sin({\frac{k (j-1)\pi}{\nu/2}}) \ket{a_j}, \\
  \mathcal{E}_k &= \mathcal{E}^{(2)}_\mathrm{d} - 2J \cos \left( {\frac{k\pi}{\nu/2}} \right) , \label{subeq:vdWeigenvalded}
  \end{align}
\end{subequations}
for $k=1,2,\ldots, \nu/2-1$. The eigenstate with the lowest energy $\mathcal{E}_1 \simeq \mathcal{E}^{(2)}_\mathrm{d} - 2J$ is $\ket{A_\nu} = \ket{\aleph_1}$, while the two decoupled and ordered AFM states $\ket{\aleph_{\nu/2,\nu/2+1}} = (\ket{a_1} \pm \ket{a_{\nu/2+1}})/\sqrt{2}$ have the highest energies $\mathcal{E}_{\nu/2,\nu/2+1} = \mathcal{E}^{(2)}_\mathrm{o}$ within this manifold of AFM-like states.  

In Fig.~\ref{fig:AFMEexpert} (lower row) we compare the spectra of the AFM ground-state manifold for different (even) number of atoms $\nu$ as obtained analytically above for $S \ll B_2$ and by the exact diagonalization of full Hamiltonian of the system $\mathcal{H} = \mathcal{H}_{\mathrm{af}} + \mathcal{H}_{\mathrm{aa}}$. We observe excellent agreement of the perturbative approach with the exact results which attests to the validity of our analysis.  

Note that for $\nu=4$ atoms considered in the main text, while $\Omega(t) \simeq \Omega_0$ and thus $S> B_2$, we have the first scenario of Eqs.~(\ref{subeq:vdWeigenval}) and 
the system follows the lowest energy state of Eq.~(\ref{eq:PXPaleph1}) having the energy $\mathcal{E}_1 \simeq -2 \Delta - (2+\sqrt{2})(S+S_B)$. 
But at the end of the pulse, as $\Omega(t) \to 0$ and $B_2 > S,S_B \to 0$, the configurations $\ket{a_1} = \ket{1r1r}$, $\ket{a_2} = \ket{r11r}$, $\ket{a_3} = \ket{r1r1}$ decouple.
Then the single lowest energy state $\ket{A_4} = \ket{\aleph_1} = \ket{a_2}$ with the energy $\mathcal{E}_1 = \mathcal{E}^{(2)}_\mathrm{d} = -2(\Delta + S + S_B)$ acquires larger population, while the remaining population is trapped, due to non-adiabatic switching off of $\Omega(t)$ and thereby $J$, in one of the states (the symmetric one) $\ket{\aleph_{2,3}} = (\ket{a_1} \pm \ket{a_3})/\sqrt{2}$ having nearly degenerate energy $\mathcal{E}_{2,3} = \mathcal{E}^{(2)}_\mathrm{o} = -2(\Delta + S) + B_2 -S_{2B} - S_B$ .

\section{Dynamical and geometric phases}

During the evolution, the state of the system $\ket{\Psi(t)}$ acquires the phase $\phi_\mathrm{tot} = \arg{\bra{\Psi(0)}\ket{\Psi(t)}}$, which is ill-defined when $\ket{\Psi(t)}$ and $\ket{\Psi(0)}$ are orthogonal, but has a well defined value when $\ket{\Psi(t)}$ and $\ket{\Psi(0)}$ have sufficient overlap. 
The total phase $\phi_\mathrm{tot} = \phi_\mathrm{d} +\phi_\mathrm{g}$ contains the dynamical $\phi_\mathrm{d}$ and geometric $\phi_\mathrm{g}$ contributions. 
The dynamical phase is given by the action $\phi_\mathrm{d} = \int_0^t \bra{\Psi(t')}\mathcal{H}(t')\ket{\Psi(t')}dt'$, and the remaining phase is geometric,  
\begin{equation}
    \phi_\mathrm{g} = \arg{\bra{\Psi(0)}\ket{\Psi(t)}} - \phi_\mathrm{d} .
\end{equation}
We rely on the symmetry of the spectrum discussed in Appendix~\ref{app:HSpctrSym} to cancel the dynamical phase during the two steps of our protocol and at the end of the process only the geometric phase remains. 

\subsection{A two-leveled atom}

To understand the nature of the geometric phase that we employ in our protocol, it is instructive to recall first the adiabatic (Landau-Zener) dynamics of a single two-level atom with levels $\ket{1}$ and $\ket{r}$ coupled by a laser with the Rabi frequency $\Omega$ and detuning $\Delta$ as described by the Hamiltonian 
\begin{equation}\label{eq:2LAFHam}
    \mathcal{H}_{\mathrm{2la}} = \tfrac{1}{2} \Omega \ket{r}\bra{1} + \mathrm{H.c.} - \Delta \ket{r}\bra{r}.  
\end{equation}
The eigenstates and the corresponding energy eigenvalues of $\mathcal{H}_{\mathrm{2la}}$ are
\begin{eqnarray}
    \ket{\alpha_{\pm}} &=& \frac{1}{\sqrt{N_{\pm}}} \left[ \left(\sqrt{\Delta^2 + |\Omega|^2} \pm \Delta \right) \ket{1} \pm \Omega \ket{r} \right] , \\
    \mathcal{E}_{\pm} &=& - \tfrac{1}{2} \left(\Delta \mp \sqrt{\Delta^2 + |\Omega|^2} \right) ,
\end{eqnarray}
where $N_\pm$ are the normalization factors. 
At $\Delta=0$, we have the minimal energy gap $|\Omega|$
between $\ket{\alpha_{\pm}} = \frac{1}{\sqrt{2}} [ \ket{1} \pm \ket{r}]$ with $\mathcal{E}_{\pm} = \pm \frac{1}{2} |\Omega|$. For large negative $\Delta$, $-\Delta \gg |\Omega|$, we have $\ket{\alpha_-} \to \ket{1}$ with $\mathcal{E}_- \to 0$ and $\ket{\alpha_+} \to \ket{r}$ with $\mathcal{E}_+ \to -\Delta >0$; 
while for large positive $\Delta \gg |\Omega|$, we have $\ket{\alpha_-} \to -\ket{r}$ with $\mathcal{E}_- \to -\Delta < 0$ and $\ket{\alpha_+} \to \ket{1}$ with $\mathcal{E}_+ \to 0$.

Hence, starting at some time $t=-\tau/2$ in $\ket{1}$, turning on $\Omega$ and sweeping the detuning $\Delta = \beta t$ with the rate $\beta = 2\Delta_0/\tau$ from a  large negative value $-\Delta_0$ to a large positive value $\Delta_0$ at $t=\tau/2$, we adiabatically follow $\ket{\alpha_-}$ and arrive at $-e^{i\phi_\mathrm{d}^{(-)}} \ket{r}$, where the dynamical phase, 
\begin{gather*}
    \phi_\mathrm{d}^{(-)} = \int_{\tau/2}^{\tau/2} \!\!\! \mathcal{E}_-(t)dt \\
    = - \frac{|\Omega| \tau}{4} \left[\sqrt{\frac{\Delta_0^2}{|\Omega|^2}+1} + \frac{|\Omega|}{\Delta_0}\sinh^{-1}\left( \frac{\Delta_0}{|\Omega|} \right) \right] ,
\end{gather*}
reduces to $\phi_\mathrm{d}^{(-)} \simeq - \tfrac{1}{4}\Delta_0 \tau$ for $\Delta_0 \gg |\Omega|$.

Repeating the same sequence again, starting now in $-e^{i\phi_\mathrm{d}^{(-)}}\ket{r}$, we adiabatically follow $\ket{\alpha_+}$ and arrive at $-e^{i\phi_\mathrm{d}^{(-)}}e^{i \phi_\mathrm{d}^{(+)}}  \ket{1} = -\ket{1}$, where the dynamical phases cancel, since $\phi_\mathrm{d}^{(+)} =  \int \mathcal{E}_+(t)dt = - \phi_\mathrm{d}^{(-)}$, and only the geometric phase $\phi_\mathrm{g}=\pi$ remains. 

Recall that at each sweep, the non-adiabatic transition probability to the other eigenstate is given by the Landau-Zener equation $P_{\mathrm{LZ}} = \exp \left( -2\pi |\Omega/2|^2/\beta \right)$ \cite{Landau1932, Zener1932}. 
Hence, good adiabatic following, $P_{\mathrm{LZ}} \ll 1$, requires either large gap $|\Omega|$ or slow sweep with a small rate $\beta$ \cite{MessiahQM1961vII}.  

Note that the sign (or phase) of $\Omega$ is a matter of convention, i.e., by changing $\Omega \to -\Omega$, we have to also exchange the eigenstates $\ket{\alpha_-} \leftrightarrow \ket{\alpha_+}$, and the sign change of the state (acquiring the geometric phase $\pi$) will not occur during the first sweep along the lower pass, but will occur during the second sweep along the upper pass. 
The crucial observation, however, is that, for any sign or phase of $\Omega$, starting from $\ket{1}$, the adiabatic  following of the lower energy eigenstate towards $\ket{r}$, followed by the adiabatic following of the higher energy eigenstate towards $\ket{1}$, leads to the transformation $\ket{1} \to -\ket{1}$.  

\subsection{Multiatom system}

The situation is completely analogous for $\nu > 1$ two-level atoms. 
If the atoms were non-interacting, each of them would proceed from $\ket{1}$ to $\ket{r}$ and back and acquire a $\pi$ phase shift, and total state would transform as $\ket{G_\nu} \to (-1)^{\nu} \ket{G_\nu}$.
In our system of strongly interacting atoms, only 
$\nu_r = \lceil \nu/2 \rceil$ atoms are excited to the Rydberg state $\ket{r}$ while the remaining atoms stay in state $\ket{1}$ due to the Rydberg blockade. Hence the total state transforms as $\ket{G_\nu} \to (-1)^{\nu_r} \ket{G_\nu}$. 
Below we present a formal proof for this intuitive statement.

\subsubsection{\textit{PXP} model}
\label{app:parityphasePXP}

During each pulse, the system is transformed according to the unitary evolution operator
\begin{equation}
\mathcal{U} \equiv 
\exp (-i \int_{-\tau/2}^{\tau/2} \mathcal{H}_{PXP}[\Omega(t), \Delta(t)]dt ) = 
\mathcal{U}[\Omega(t), \Delta(t)] \label{eq:UtrPXP}
\end{equation}
that does not depend on the strength of the interatomic interactions $B$ taken care of by the structure of the $PXP$ model.
Acting with $\mathcal{U}$ on state $\ket{G_\nu}$, we have 
\begin{equation}
    \mathcal{U}\ket{G_\nu} = \ket{A_\nu} , \label{eq:UGtoA}
\end{equation}
were the total phase is included in the definition of the AFM state $\ket{A_\nu}$. 
Assuming the laser pulse with time-symmetric $\Omega(t) = \Omega(-t)$ (even function) and anti-symmetric $\Delta(t) = - \Delta(-t)$ (odd function), for the inverse unitary operator, $\mathcal{U}^{-1} \mathcal{U} = \mathcal{U}\mathcal{U}^{-1} = \mathbb{1}$, we have 
\begin{align*}
\mathcal{U}^{-1} & \equiv 
\exp (i \int_{\tau/2}^{-\tau/2} \mathcal{H}_{PXP}[\Omega(t), \Delta(t)] dt) \\
& = \exp (-i \int_{-\tau/2}^{\tau/2} \mathcal{H}_{PXP}[-\Omega(t), \Delta(t)]dt ) \\
& = \mathcal{U}[-\Omega(t), \Delta(t)] .
\end{align*}

Next, we use the parity operator 
\begin{equation}
    \mathcal{P} = \prod_{i=1}^\nu ( \ket{1_i}\bra{1_i} - \ket{r_i}\bra{r_i}) , \label{eq:Parity}
\end{equation}
which is its own inverse, $\mathcal{P} \mathcal{P} = \mathbb{1}$, and acts as 
\begin{subequations}
    \begin{align}
        \mathcal{P}\ket{G_\nu} &= \ket{G_\nu} , \\
        \mathcal{P}\ket{A_\nu} &= (-1)^{\nu_r}\ket{A_\nu} .
    \end{align}
\end{subequations}
This operator anticommutes with the first term ($\propto \Omega$) of the Hamiltonian $\mathcal{H}_{PXP}$ in Eq.~(\ref{eqs:PXPhamiltonian}) and commutes with the second term ($\propto \Delta$), which implies that 
$\mathcal{P}\mathcal{U}[\Omega(t), \Delta(t)] = \mathcal{U}[-\Omega(t), \Delta(t)]\mathcal{P}$ or equivalently 
\begin{equation}\label{eq:UPandPU}
    \mathcal{P}\mathcal{U} = \mathcal{U}^{-1}\mathcal{P} , \quad  \mathcal{P}\mathcal{U}^{-1} = \mathcal{U}\mathcal{P}.
\end{equation}

Starting with $\ket{G_\nu} = \mathcal{U}^{-1} \mathcal{U} \ket{G_\nu}$ and acting on both sides by $\mathcal{P}$, we can now write 
\begin{align*}
\ket{G_\nu} &= \mathcal{P} \mathcal{U}^{-1} \mathcal{U} \ket{G_\nu} = \mathcal{U} \mathcal{P} \ket{A_\nu} \\
&= (-1)^{\nu_r} \mathcal{U} \ket{A_\nu} = (-1)^{\nu_r} \mathcal{U}^2 \ket{G_\nu} ,
\end{align*}
and therefore 
\begin{equation}\label{eq:UUG}
    \mathcal{U}^2\ket{G_\nu}  = (-1)^{\nu_r}\ket{G_\nu} , 
\end{equation}
which means that consecutive application of two identical pulses returns the system to the ground state $\ket{G_\nu}$ and accords the geometric phase $\phi_\mathrm{g}= \nu_r \pi \! \mod(2\pi)$.

\subsubsection{van der Waals interacting model}
\label{app:parityphaseVdW}

Consider now the full model described by the Hamiltonian $\mathcal{H} = \mathcal{H}_\mathrm{af} + \mathcal{H}_\mathrm{aa}$ that includes the interatomic interactions.
The unitary evolution operator 
\begin{align}
    \mathcal{U} &\equiv
    \exp (-i \int_{-\tau/2}^{\tau/2} \mathcal{H}[\Omega(t), \Delta(t), B]dt ) \nonumber \\
    &\equiv \mathcal{U}[\Omega(t), \Delta(t), B] \label{eq:UtrVdW}
\end{align} 
is now a function of interaction $B$.
Our goal is to show that $\mathcal{U}_\mathrm{II} \mathcal{U}_\mathrm{I} \ket{G_\nu}  = (-1)^{\nu_r}\ket{G_\nu}$ where $\mathcal{U}_{\mathrm{I}} = \mathcal{U}[\Omega(t), \Delta(t), B]$ and $\mathcal{U}_{\mathrm{II}} = \mathcal{U}[\Omega(t), \Delta(t), -B]$ describe the unitary evolution during Steps I and II.

Starting in state $\ket{G_\nu}$ and assuming adiabaticity, the action of $\mathcal{U} = \mathcal{U}_\mathrm{I}$ results in $\mathcal{U}_\mathrm{I} \ket{G_\nu} = \ket{A_\nu}$, while for the inverse unitary transformation we have 
\begin{align*}
\mathcal{U}^{-1} & \equiv \exp (i \int_{\tau/2}^{-\tau/2} \mathcal{H}[\Omega(t), \Delta(t), B] dt) \\
& = \exp (-i \int_{-\tau/2}^{\tau/2} \mathcal{H}[-\Omega(t), \Delta(t), -B]dt ) \\
& = \mathcal{U}[-\Omega(t), \Delta(t), -B] ,
\end{align*}
where we assumed, as before,  $\Omega(t) = \Omega(-t)$ (even function of time) and $\Delta(t) = - \Delta(-t)$ (odd function). 
The parity operator of Eq.~(\ref{eq:Parity}) anticommutes with the term $\propto \Omega$ of the Hamiltonian $\mathcal{H}$, while commutes with the terms $\propto \Delta$ and $\propto B$. 
This implies that $\mathcal{P}\mathcal{U}[\Omega(t), \Delta(t), B] = \mathcal{U}[-\Omega(t), \Delta(t), B]\mathcal{P}$ and thus
\begin{equation}\label{eq:PUIandUIP}
    \mathcal{P}\mathcal{U}_\mathrm{I}^{-1} = \mathcal{U}_\mathrm{II}\mathcal{P}, \qquad \mathcal{U}_\mathrm{I}^{-1} \mathcal{P}= \mathcal{P}\mathcal{U}_\mathrm{II}
\end{equation}
 
Starting from the identity $\ket{G_\nu} = \mathcal{U}_\mathrm{I}^{-1} \mathcal{U}_\mathrm{I} \ket{G_\nu}$ and acting on both sides by $\mathcal{P}$, we can now write 
\begin{align*}
\ket{G_\nu} &= \mathcal{P} \mathcal{U}_\mathrm{I} ^{-1} \mathcal{U}_\mathrm{I}  \ket{G_\nu} = \mathcal{U}_\mathrm{II}  \mathcal{P} \ket{A_\nu} \\
&= (-1)^{\nu_r} \mathcal{U}_\mathrm{II}  \ket{A_\nu} = (-1)^{\nu_r} \mathcal{U}_\mathrm{II} \mathcal{U}_\mathrm{I} \ket{G_\nu} ,
\end{align*}
and therefore 
\begin{equation}\label{eq:UIIUIG}
    \mathcal{U}_\mathrm{II} \mathcal{U}_\mathrm{I}\ket{G_\nu}  = (-1)^{\nu_r}\ket{G_\nu} , 
\end{equation}
which means that consecutive application of two identical pulses, while flipping the sign of interaction $B$ between the pulses, returns the system to the ground state $\ket{G_\nu}$ and assigns the geometric phase $\phi_g= \nu_r \pi \! \mod(2\pi)$.

\section{Error due to transfer to states with wrong parity}
\label{app:Eleakagewp}

So far we have assumed that the evolution operator $\mathcal{U}$ in Eq.~(\ref{eq:UtrPXP}) or (\ref{eq:UtrVdW}) results in transformation (\ref{eq:UGtoA}) where only the state $\ket{A_\nu}$ with the correct party is populated at the end of the first pulse $t=\tau$. 
In reality, non-adiabatic transitions away from the ground state can lead to populations of states other than $\ket{A_\nu}$, but only the states with the wrong party lead to gate errors. 
Let us assume that 
\begin{equation}
\mathcal{U}\ket{G_\nu} = a \ket{A_\nu} + b \ket{B_\nu} , \label{eq:UGaAbB}
\end{equation}
where $\ket{A_\nu}$ is the target state with parity $\mathcal{P}\ket{A_\nu} = (-1)^{\nu_r} \ket{A_\nu}$ as before, while $\ket{B_\nu}$ is the state with the wrong parity $\mathcal{P}\ket{B_\nu} = (-1)^{\nu_r \pm 1} \ket{B_\nu}$, and their amplitudes are normalized as $|a|^2 + |b|^2 = 1$ with $|a|^2 \gg |b|^2$. 
Starting with the identity $\ket{G_\nu} = \mathcal{U}^{-1} \mathcal{U} \ket{G_\nu}$ and proceeding as before, we have 
\begin{align*}
\ket{G_\nu} &= \mathcal{P} \mathcal{U}^{-1} \mathcal{U} \ket{G_\nu} = \mathcal{U}\mathcal{P}[ a \ket{A_\nu} + b \ket{B_\nu}] \\
&= (-1)^{\nu_r}\mathcal{U}[ a \ket{A_\nu} - b \ket{B_\nu}]  \\
&= (-1)^{\nu_r}[\mathcal{U}^2 \ket{G_\nu} -2 b \, \mathcal{U}\ket{B_\nu}] ,
\end{align*}
where in the last line we used Eq.~(\ref{eq:UGaAbB}) to express $a\ket{A_{\nu}}$. 
At time $\tau_{\mathrm{tot}}$, the amplitude $g(\tau_{\mathrm{tot}}) = \bra{G_\nu} \mathcal{U}^2\ket{G_\nu}$ of the initial state $\ket{G_\nu}$ is then 
\begin{align}
g(\tau_{\mathrm{tot}}) &= (-1)^{\nu_r} + 2b\bra{G_\nu}\mathcal{U}\ket{B_\nu} \nonumber\\
&= (-1)^{\nu_r} + 2b\bra{G_\nu}\mathcal{U}\mathcal{P}^2\ket{B_\nu} \nonumber\\
&= (-1)^{\nu_r} + 2b\bra{G_\nu}\mathcal{U}^{-1}\mathcal{P}\ket{B_\nu}\nonumber\\
&= (-1)^{\nu_r} [1 - 2|b|^2] ,
\end{align}
and the error probability due to leakage to state $\ket{B_\nu}$ is $E_{\mathrm{leak}} = 1-|g|^2 \simeq 4|b|^2$.

\section{Changing the sign of interaction \textit{B}}
\label{app:BtomB}

%%%%%%%%%%%%%%%%%%%%%%%%%%%%%%%%%%%%%%%%%%%%%%%%
\begin{figure}[h!]
\includegraphics[width=\columnwidth]{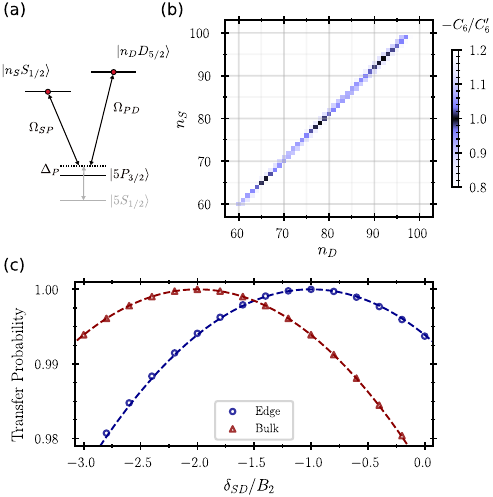}
\caption{(a) Relevant levels and transitions of Rb atoms for a two-photon transfer between the Rydberg states $\ket{r} = \ket{n_S S_{1/2}}$ and $\ket{r'} =\ket{n_D D_{5/2}}$ via the intermediate non-resonant state $\ket{5P_{3/2}}$ (or $\ket{6P_{3/2}}$). 
(b) States $\ket{n_S S_{1/2}}$ have positive vdW coefficient, $C_6 > 0$, while states $\ket{n_D D_{5/2}}$ have negative vdW coefficient, $C_6' < 0$ (assuming small or vanishing electric and magnetic fields) \cite{Singer_2005}. With a proper choice of the principal quantum numbers $n_S$ and $n_D$, the ratio $-C_6/C_6'$ can be close to 1. 
(c) Probability of transfer $\ket{r} \to \ket{r'}$ for an edge atom (red open triangles) or a bulk atom (blue open circles) by a two-photon $\pi$-pulse $\int |\Omega_{SD}| dt = \pi$ vs the two-photon detuning $\delta_{SD}$, as obtained from numerical simulations of the transfer dynamics in a chain of atoms initially in AFM-like state subject to a square pulse with $|\Omega_{SD}|/2\pi = 8\:$MHz and duration $t_{\mathrm{tr}}=\pi/|\Omega_{SD}|$, while $B \simeq -B' \simeq 2\pi \times 45\:$MHz ($B_2=B/64$). 
Dashed lines of the same color correspond to the mean-field expressions (\ref{eq:EtrBE}).}
\label{fig:TrBmB}
\end{figure}
%%%%%%%%%%%%%%%%%%%%%%%%%%%%%%%%%%%%%%%%%%%%%%%%

We assume Rb atoms excited from the ground state sublevel $\ket{1} = \ket{5S_{1/2},F=2,M_F=0}$ to the Rydberg state $\ket{r} = \ket{n_S S_{1/2}}$ 
(and de-excited from the Rydberg state $\ket{r'} = \ket{n_D D_{5/2}}$) 
via a two-photon transition involving an intermediate non-resonant state $\ket{5P_{3/2}}$ (or $\ket{6P_{3/2}}$), as is done in most experiments \cite{evered2023high,graham2019,graham2022,Levine2019Parallel} and illustrated in Fig.~\ref{fig:TrBmB}(a).
In weak or vanishing electric and magnetic fields, van der Waals interaction between the atoms in states $\ket{n S_{1/2}}$ is repulsive, $C_6 > 0$ ($B>0$), 
while in states $\ket{n D_{5/2}}$ is attractive $C_6' < 0$ ($B' <0$).
With a proper choice of the principal quantum numbers $n_S$ and $n_D$ for the $\ket{n_S S_{1/2}}$ and $\ket{n_D D_{5/2}}$ Rydberg states, their interactions can, with high precision, be $B\simeq -B'$ ($C_6 \simeq -C_6'$) \cite{Singer_2005}, see Fig.~\ref{fig:TrBmB}(b).  

To change the sign of the interaction, we can thus transfer the atoms from state $\ket{r} = \ket{n_S S_{1/2}}$ to state $\ket{r'} = \ket{n_D D_{5/2}}$ via a two-photon (Raman) process through the intermediate non-resonant state $\ket{5P_{3/2}}$ (or $\ket{6P_{3/2}}$) using a pair of laser fields with Rabi frequencies $\Omega_{SP}$ and $\Omega_{PD}$ and detuning $\Delta_P \gg \Omega_{SP,PD}$, see Fig.~\ref{fig:TrBmB}(a). 
By using optical transitions for the transfer, we avoid resonant dipole-dipole interactions between the atoms on the dipole-allowed Rydberg transitions with large dipole moments.

Consider a chain of atoms in an AFM-like configuration of Rydberg excitations, $\ket{r1r1r\ldots}$. 
To fully transfer the atoms from state $\ket{r}$ to state $\ket{r'}$, the effective two-photon Rabi frequency for the transition $\ket{r} \to \ket{r'}$, $\Omega_{SD} = \Omega_{SP} \Omega_{PD}/2\Delta_P$, should be much larger than the interaction strengths $B_2=B/2^6$ and $B_2'=B'/2^6$ between the next-nearest-neighbor atoms, $|\Omega_{SD}| \gg |B_2^{(\prime)}|$. 
Simultaneously, the two-photon detuning $\delta_{SD}$, containing also the ac Stark shifts $|\Omega_{SP,PD}|^2/{4\Delta_P}$ of levels $\ket{r,r'}$, should be chosen such as to partially or fully compensate the interaction-induced level shifts of the atoms.
For an atom in the AFM bulk and having two next-nearest neighbors, the frequency of transition $\ket{r} \to \ket{r'}$ is initially shifted by $2B_2$ due to the level shift of $\ket{r}$, is shifted by $-2B_2'$ at the end of the transfer due to the level shift of $\ket{r'}$, and is shifted by about $(B_2-B_2')$ during the transfer when the atoms are in the superposition of states $\ket{r}$ and $\ket{r'}$, and for simplicity we neglect the interaction between the atoms in states $\ket{r}$ and $\ket{r'}$. 
For a Rydberg atom at the edge of the chain, or near the defect of the AFM-like configuration, and therefore having only one next-nearest neighbor, the same arguments as above lead to the mean shift of the $\ket{r} \to \ket{r'}$ transition frequency being $(B_2 - B_2')/2$. 
Assuming a $\pi$-pulse, $\int |\Omega_{SD}| dt = \pi$, with some two-photon detuning $\delta_{SD}$, we thus estimate that the transfer errors for an atom in the bulk or at the edge is
\begin{subequations}
\label{eq:EtrBE}
\begin{eqnarray}
E_{\mathrm{bulk}} &\simeq& \left| \frac{\delta_{SD} + (B_2-B_2')}{\Omega_{SD}}\right|^2 , \\
E_{\mathrm{edge}} &\simeq& \left| \frac{\delta_{SD} + \frac{1}{2} (B_2-B_2')}{\Omega_{SD}}\right|^2 .
\end{eqnarray}
\end{subequations}
We also assume that the phase of the two-photon pulse is $\arg({\Omega_{SD}) = \pi/2}$ so that the transfer $\ket{r} \to \ket{r'}$ does not introduce a sign change to the Rydberg state of each atom.

We verified these arguments by performing numerical simulations of the $\ket{r} \to \ket{r'}$ transfer for $N \geq 5$ atoms initially in the antiferromagnetic-like state of Rydberg excitations without or with defects in the bulk, e.g., $\ket{r1r1r\ldots}$ or $\ket{r11r1\ldots}$. Upon applying the two-photon $\pi$-pulse to all the atoms, the resulting probabilities of transfer to state $\ket{r_j'}$ for the atom $j$ in the bulk or the edge, $\braket{\Psi}{ r_j'} \braket{r_j'}{\Psi}$, are shown in Fig.~\ref{fig:TrBmB}(c) as a function of two-photon detuning $\delta_{SD}$. We observe excellent agreement between the analytic mean-field expressions (\ref{eq:EtrBE}) and exact numerical results. 

Hence, if we set the two-photon detuning $\delta_{SD} \simeq -(B_2-B_2') \simeq -2B_2$ to compensate the interaction-induced level shifts of the atoms in the AFM bulk, $E_{\mathrm{bulk}} \simeq 0$, the transfer error for the edge atoms, or the atoms adjacent to the defect in the AFM-like configuration, will be $E_{\mathrm{edge}}\simeq \left| \frac{B_2}{\Omega_{SD}} \right|^2$. 
More generally, for a configuration with $\nu_{\mathrm{edge}}$ edge atoms and $\nu_{\mathrm{bulk}}$ atoms, the total transfer error is 
$E_{\mathrm{transfer}} \simeq \nu_{\mathrm{edge}} E_{\mathrm{edge}} + \nu_{\mathrm{bulk}} E_{\mathrm{bulk}}$. 
With experimentally relevant parameters, $B_2 \simeq -B_2' \ll |\Omega_{SD}|$, we obtain the minimal transfer error per bulk atom to be $ E_{\mathrm{bulk}}<10^{-6}$ and simultaneously per edge atom to be $E_{\mathrm{edge}}<10^{-2}$ [or the other way around, depending on $\delta_{SD}$; see Fig.~\ref{fig:TrBmB}(c)]. 
The total transfer error $E_{\mathrm{transfer}} \lesssim 0.01$ is then smaller than other sources of error.  
With amplitude- and phase-optimized pulses, the transfer error can be significantly reduced even further. 

%%%%%%%%%%%%%%%%%%%%%%%%%%%%%%%%%

\bibliography{refs}

\end{document}